%% file: manuscript_jhep.tex
\documentclass[11pt,a4paper]{article}
\usepackage{jheppub}
\usepackage[T1]{fontenc}
\usepackage[utf8]{inputenc}
\usepackage{lmodern}
\usepackage{microtype}
\usepackage{bm,booktabs,array,tabularx,xcolor,tikz}
\usetikzlibrary{decorations.pathmorphing,decorations.markings,arrows.meta}
\newcommand{\dd}{\mathrm d}
\newcommand{\eps}{\epsilon}
\newcommand{\UV}{\mathrm{UV}}
\newcommand{\IR}{\mathrm{IR}}
\newcommand{\intq}{\int\!\frac{\dd^dq}{(2\pi)^d}}
\newcommand{\intp}{\int\!\frac{\dd^dp}{(2\pi)^d}}
\hypersetup{pdftitle={Exact two-loop gravitational tail integrals and their ultraviolet-infrared structure},
 pdfauthor={Gabriel Luz Almeida}}
\title{Exact two-loop gravitational tail integrals and their ultraviolet-infrared structure}
\author{Gabriel Luz Almeida}
\affiliation{School of Physics \& Shing-Tung Yau Center, Southeast University,\\
Nanjing 210096, China}
\emailAdd{galmeida@seu.edu.cn}

\abstract{
Gravitational tail-of-tail corrections arise when radiation scatters twice from the
mass-monopole field of its source. In the multipolar effective field theory, one
of the two-loop topologies contributing to this process involves genuinely coupled loop momenta, preventing its evaluation through nested one-loop integrations. After tensor reduction,
this nonfactorizing structure leads to a particularly challenging family of
scalar integrals with unbounded numerator powers as the multipole rank
increases. We derive exact closed-form expressions for the entire relevant
integer numerator domain, valid in generic dimension and accommodating
Feynman, retarded, and advanced prescriptions through consistent analytic
continuation. Near $d=3+\epsilon$, we also obtain explicit Laurent expansions
through $O(\epsilon)$, retaining the subleading information needed in
higher-order radiative calculations. While these expansions determine the
dimensional poles, identifying their ultraviolet (UV) or infrared (IR) origin
requires a separate analysis of momentum regions. We carry out this analysis
throughout the integral domain and apply it directly to the coupled tail-of-tail
emission diagram for arbitrary electric and magnetic multipoles. In this way, we
independently determine the UV and IR residues, and explicitly verify that the
complete infrared pole structure, including both double and simple poles,
agrees with the exponentiation of the one-loop gravitational tail. The UV
residues, calculated directly from hard-momentum regions, reproduce the
established multipole anomalous dimensions when combined with the known
contributions of the remaining diagrams.
}

\begin{document}
\maketitle
\flushbottom

\section{Introduction}
\label{sec:intro}

Gravitational waves scatter from the long-range field of the source that
emits them. One interaction with its mass-monopole field produces a tail,
while two produce a tail of a tail, introducing delayed, history-dependent
contributions and logarithms into the waveform. These nonlinear
propagation effects are systematically described within the
multipolar-post-Minkowskian framework \cite{BlanchetDamour1986}, where
hereditary contributions were identified through their dependence on
the source's past history \cite{BlanchetDamour1992}. In particular, the
cubic interaction responsible for gravitational tails of tails was
calculated explicitly in ref.~\cite{Blanchet1998}.

The same propagation effects can be described within the
long-wavelength effective field theory (EFT) of gravitational
radiation, which separates the internal dynamics of a compact
source from the propagation of its emitted waves
\cite{GoldbergerRothstein2006}. In this framework, an effective
multipolar action organizes the source couplings to radiation
in a derivative expansion \cite{Ross2012}. Tail-of-tail
corrections are then represented by emission diagrams involving
two interactions of the radiative field with the source's
long-range mass-monopole field. Using this approach, Goldberger
and Ross calculated the quadrupolar tail-of-tail correction
and showed that its ultraviolet (UV) divergences renormalize
the source multipole, giving rise to renormalization-group
logarithms \cite{GR2010}. The subsequent extension to arbitrary
electric and magnetic multipoles \cite{AFS2021} further
established the interplay between short-distance renormalization
and the infrared (IR) singularities associated with long-range
wave propagation.

The significance of these radiative effects extends beyond the
emission amplitudes themselves. When radiation is reabsorbed by the
source, hereditary interactions also enter its dynamics through
nonlocal contributions to the effective action, connecting radiation
reaction with the matching of near- and far-zone descriptions
\cite{FS2013,Galley2016}. Related loop calculations arise in
conservative tail interactions
\cite{AFSConservative2021,FSHereditary2020}, while nonlinear
interactions among radiative fields generate gravitational memory
\cite{AMFSMemory2025,Porto2025}. Although these processes contribute
to different observables, they share analytical challenges associated
with multiloop integration and the treatment of long-distance
singularities.

These challenges become increasingly relevant at higher orders in
gravitational radiation. Generalized unitarity and diagrammatic
methods have been used to organize higher-order quadrupolar tails
\cite{EdisonLevi2023,Edison2025}, while the universal logarithmic
contributions have been studied through renormalization-group
resummation \cite{Ivanov2025}. Repeated mass-tail interactions also
dress gravitational memory with soft logarithms \cite{Akhtar2026}.
Such calculations require increasingly detailed information about
the underlying integrals. In particular, when lower-order
contributions multiply additional singular factors, coefficients
beyond the leading poles and finite terms can become essential.

The mathematical difficulty is already apparent at the tail-of-tail
level. Of the three two-loop emission topologies associated with two
scatterings of multipolar radiation off the source's mass-monopole
field, two admit sequential integrations in terms of standard
one-loop formulas. The remaining topology is
considerably more involved: its radiative propagators are linked
by static exchanges, coupling the loop momenta in a way that prevents
a straightforward reduction to nested one-loop integrations.
This difficulty was encountered in the original quadrupolar
calculation of Goldberger and Ross \cite{GR2010}, who employed
Mellin--Barnes representations to evaluate the relevant coupled
integrals. Although the required results could ultimately be
expressed in Gamma functions, their calculation illustrates the
nontrivial analytical structure of this topology.

For arbitrary multipoles, the problem becomes more demanding.
As the multipole rank increases, the multipolar source couplings
contain additional spatial derivatives, leading to higher powers
of loop momenta in the tensor numerators. After tensor reduction, these
contributions generate scalar integrals with numerator powers
of arbitrarily high degree. Thus, rather than evaluating a finite
set of integrals associated with a particular multipole, a general
treatment requires solving an entire integer domain of propagator
indices. The main objective of this work is to obtain an exact
analytical solution for that domain, without repeating the
integration for each multipole rank.

Integration-by-parts (IBP) identities provide a natural starting
point for this problem. By relating integrals whose propagator
indices differ by integer shifts \cite{Tkachov1981,ChetyrkinTkachov1981},
they underlie systematic reduction methods such as the Laporta
algorithm \cite{Laporta2000} and its automated implementations,
including FIRE \cite{SmirnovChuharev2020}. While these methods
efficiently reduce integrals at fixed indices, the present problem
calls for analytical control over an unbounded set of numerator
powers. Related approaches exploit dimensional recurrences
\cite{Tarasov1996}, dimensional analyticity \cite{Lee2010}, or
differential equations in external invariants \cite{Henn2013};
comprehensive discussions can be found in
refs.~\cite{Smirnov2012,Weinzierl2022}. The growing importance
of such techniques in classical gravity is illustrated, for
example, by multiloop calculations of post-Minkowskian tail
contributions \cite{Bern2022}.

For the on-shell family considered here, the solution takes a
particularly compact form. The relevant integer domain naturally
decomposes into four sectors, distinguished by the propagators
that remain after numerator reduction. Using three IBP identities
together with boundary integrals expressible through standard
one-loop formulas, we solve the corresponding recurrences exactly.
The resulting expressions are given entirely in Gamma functions
and cover the full required integer numerator domain in generic
dimension. Moreover, while the algebraic recurrences do not
depend on the causal prescription, their Feynman, retarded and
advanced boundary values are obtained through consistent analytic
continuation of the on-shell invariant and the boundary integrals.
We also derive explicit Laurent expansions through $O(\eps)$ near
three spatial dimensions. The subleading coefficients are relevant
for higher-order radiative calculations, including further
iterations of the tail process, such as tail-of-tail-of-tail
interactions, where lower-order contributions enter multiplied
by additional singular factors.

Although these exact expressions determine the dimensional poles,
they do not by themselves identify their physical origin.
Dimensional regularization can conceal this distinction because
scaleless integrals vanish even when they contain compensating UV
and IR contributions. Consequently, reducing an integral to a
set of simpler expressions does not necessarily preserve the
information about which momentum regions generate its poles.
Separating these contributions requires identifying the relevant
short- and long-distance regions and accounting consistently for
their overlaps. In EFT, this issue is addressed by the zero-bin
subtraction \cite{ManoharStewart2007}, whose application to
gravitational near/far-zone matching has clarified the assignment
of UV and IR singularities
\cite{PortoRothstein2017,FPRS2019,FS2021}. More generally,
expansion by regions provides a systematic way to isolate the
contributions associated with different momentum scalings
\cite{BenekeSmirnov1998,Jantzen2011}. Another related framework
is the $R^*$ operation, developed to organize UV and IR
counterterm subtractions within dimensional regularization
\cite{ChetyrkinTkachov1982,ChetyrkinSmirnov1984}, and subsequently
extended to more general numerator structures
\cite{HerzogRuijl2017}. In the present work, we determine the
pole origins directly through a momentum-region analysis of the
integrals. This yields a complete UV/IR classification of the
four sectors near $d=3+\eps$, including two sectors whose exact
regulated values coincide despite their poles having opposite
momentum origins.

The same region analysis allows us to determine the UV and IR
contributions of the coupled gravitational emission diagram for
electric and magnetic multipoles of arbitrary rank. We use the
conventions, Feynman rules and unintegrated amplitudes of
ref.~\cite{AFS2021}, while evaluating their singular contributions
independently in the hard- and soft-momentum regions. This
provides a direct test of gravitational IR exponentiation, which
organizes the singular effects of repeated long-range
interactions in terms of lower-order soft data
\cite{Weinberg1965,NaculichSchnitzer2011}. In particular, we
verify that both the double and simple IR poles agree with the
exponentiation of the one-loop tail, including the contribution
from its finite part to the simple pole. The UV residues are
determined independently rather than inferred by subtracting
the expected IR contribution. When combined with the known
UV contributions of the two topologies reducible to one-loop
integrations, they reproduce the established common multipole
anomalous dimension.

Beyond the mass-monopole problem, these exact results also
provide analytical ingredients for tail-of-tail interactions
involving angular momentum. Angular-momentum couplings have
already been studied in failed-tail emission and conservative
processes \cite{AMFS2023,AMFS2026}, motivating their extension
to higher-order tail interactions. Replacing a static
mass-monopole insertion by an angular-momentum coupling
preserves the propagator structure of the corresponding
tail-of-tail topologies while modifying their tensor
numerators. In particular, the coupled successive-scattering
topology leads to the same scalar integral family solved here,
so that its analytical integration requires no new set of
master integrals. Another natural direction is the extension
to theories beyond general relativity, where tail-of-tail effects 
remain to be systematically investigated. Such
calculations would build on the existing EFT description
of scalar-tensor radiation \cite{KPV2019} and previous
studies of tails and hereditary effects
\cite{BBT2022,Trestini2025}. Although additional degrees of
freedom may introduce new propagator structures, the exact
results obtained here remain directly applicable to
contributions sharing the same scalar denominator family.

In this paper, we begin by introducing the gravitational
tail-of-tail topologies and the associated scalar integral
family in sections~\ref{sec:eft} and \ref{sec:family}.
We then develop the IBP reduction strategy and obtain the
exact analytical solution in sections~\ref{sec:ibp} and
\ref{sec:exact}. With these results established, we turn
to the origin of the dimensional poles in
section~\ref{sec:poles}, before applying the UV/IR
separation to gravitational emission amplitudes,
infrared exponentiation and multipole renormalization
in section~\ref{sec:amplitudes}. The appendices
provide the one-loop boundary integrals, detailed
recurrences, dimensional expansions and explicit
hard-region calculations supporting the main results.

\section{Tail-of-tail integrals in gravitational EFT}
\label{sec:eft}

In the long-wavelength effective theory of gravitational radiation, the
internal dynamics of a compact source is encoded in its multipole moments,
while the propagation of the emitted waves is treated explicitly
\cite{GoldbergerRothstein2006,Ross2012,GR2010,AFS2021}.
At distances large compared with the source size, these time-dependent
multipoles describe its radiative structure, whereas the total
mass-energy $E$ sources the long-range gravitational field through
which the radiation propagates. The tail-of-tail correction arises
at second order in the interaction of the emitted radiation with
this field. We first examine the corresponding two-loop emission
topologies, before identifying the scalar integrals needed to
treat radiating multipoles of arbitrary rank.

\subsection{Three emission topologies}

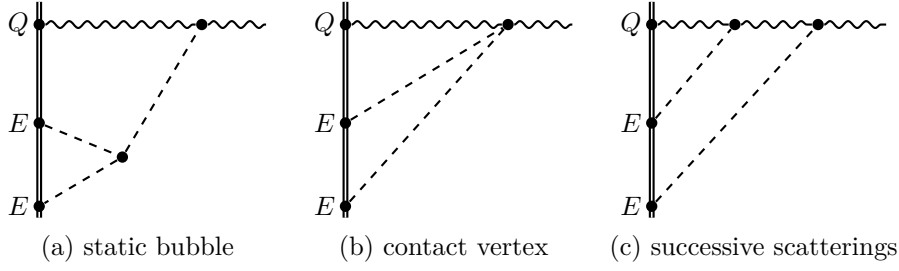
\begin{figure}[t]
\centering
\begin{tikzpicture}[x=1cm,y=1cm,
 wave/.style={decorate,decoration={snake,amplitude=.45mm,segment length=2.4mm},thick},
 potential/.style={dashed,thick},
 dot/.style={circle,fill,inner sep=1.5pt},
 every node/.style={font=\small}]
 \begin{scope}
  \draw[double,thick] (0,-.15)--(0,2.7);
  \node[dot] (aQ) at (0,2.4) {};
  \node[dot] (aM) at (0,1.1) {};
  \node[dot] (aN) at (0,0) {};
  \node[left] at (aQ) {$Q$};
  \node[left] at (aM) {$E$}; \node[left] at (aN) {$E$};
  \node[dot] (av) at (2.15,2.4) {};
  \node[dot] (ab) at (1.1,.65) {};
  \draw[wave] (aQ)--(av);
  \draw[wave] (av)--(3,2.4);
  \draw[potential] (aM)--(ab)--(aN);
  \draw[potential] (ab)--(av);
  \node at (1.3,-.55) {(a) static bubble};
 \end{scope}
 \begin{scope}[xshift=4.05cm]
  \draw[double,thick] (0,-.15)--(0,2.7);
  \node[dot] (bQ) at (0,2.4) {};
  \node[dot] (bM) at (0,1.1) {};
  \node[dot] (bN) at (0,0) {};
  \node[left] at (bQ) {$Q$};
  \node[left] at (bM) {$E$}; \node[left] at (bN) {$E$};
  \node[dot] (bv) at (2.15,2.4) {};
  \draw[wave] (bQ)--(bv);
  \draw[wave] (bv)--(3,2.4);
  \draw[potential] (bM)--(bv)--(bN);
  \node at (1.3,-.55) {(b) contact vertex};
 \end{scope}
 \begin{scope}[xshift=8.1cm]
  \draw[double,thick] (0,-.15)--(0,2.7);
  \node[dot] (cQ) at (0,2.4) {};
  \node[dot] (cM) at (0,1.1) {};
  \node[dot] (cN) at (0,0) {};
  \node[left] at (cQ) {$Q$};
  \node[left] at (cM) {$E$}; \node[left] at (cN) {$E$};
  \node[dot] (cv) at (1.1,2.4) {};
  \node[dot] (cw) at (2.2,2.4) {};
  \draw[wave] (cQ)--(cv);
  \draw[wave] (cv)--(cw);
  \draw[wave] (cw)--(3.1,2.4);
  \draw[potential] (cM)--(cv);
  \draw[potential] (cN)--(cw);
  \node at (1.35,-.55) {(c) successive scatterings};
 \end{scope}
\end{tikzpicture}
\caption{The three mass-monopole tail-of-tail emission topologies.
 The double line represents the composite source, with $Q$
 denoting a generic multipolar coupling and $E$ a static
 mass-monopole insertion. Wavy lines carry frequency $\omega$,
 while dashed lines carry zero frequency.}
\label{fig:topologies}
\end{figure}

In the EFT description, the tail-of-tail process receives
contributions from the three emission topologies shown in
figure~\ref{fig:topologies}, labelled (a), (b) and (c) following
refs.~\cite{GR2010,AFS2021}. Because the interactions with the
static monopole transfer no energy, the radiative lines retain
the emitted frequency $\omega$ throughout the diagrams. Energy
conservation thus fixes the internal frequencies, leaving two
integrations over spatial loop momenta. The distinction between
the three topologies lies in how these momenta enter the
propagators.

Although all three diagrams involve nontrivial tensor numerators,
their loop integrations differ considerably in complexity. In
diagram (a), the static lines form a massless bubble whose
integration, including the required polynomial numerator moments,
leaves a standard one-loop integral with dimension-dependent
propagator powers. Diagram (b) admits a similar treatment: the
static subintegration can be performed independently before
evaluating the remaining loop. Both topologies therefore reduce
to successive applications of known one-loop formulas, even
when the numerator structures become more involved.

The situation is different for diagram (c), where two radiative
propagators are connected by a static exchange, with a second
static exchange linking the interaction to the source. The
resulting dependence on the two loop momenta prevents the
same straightforward sequence of one-loop integrations.
This genuinely coupled topology is the central integration
problem of the present work. As we shall see, its solution
can nevertheless be constructed from simpler integrals obtained
by canceling selected propagators, which provide the boundary
data for the IBP recurrences.

\subsection{Why arbitrary multipoles require an integer-domain solution}

The difficulty of the coupled topology grows with the rank
of the radiating multipole. Higher-rank multipolar source
couplings involve additional spatial derivatives, which become
powers of momenta in Fourier space. Combined with the momentum
dependence of the bulk interactions, they generate increasingly
high-rank tensor numerators involving the independent loop
momenta $q$ and $p$ and the external momentum $k$. After
tensor reduction and contraction of the indices, these
numerators give rise to polynomials in the available scalar
products of momenta.

For diagram (c), four physical inverse propagators, together
with one additional quadratic variable representing an
irreducible scalar product, provide a complete parametrization
of these scalar products. Expressing the numerator polynomials
in terms of these variables lowers the propagator indices and
may turn them negative. Since the numerator degree grows
without bound with the multipole rank, the scalar integrals
must be evaluated over an unbounded domain of integer indices,
rather than for a finite set of fixed multipoles.

Solving this domain analytically is the central objective of
the following sections. Throughout the paper, a \emph{scalar
Feynman integral} denotes an integral with no free tensor
indices, obtained after tensor reduction. Although motivated
by the gravitational tail-of-tail problem, the resulting
solutions are properties of the integral family rather than
of a particular emission amplitude. As a concrete application,
section~\ref{sec:amplitudes} combines these results with the
unintegrated amplitudes of ref.~\cite{AFS2021} and a direct
momentum-region analysis to identify the UV and IR contributions
for arbitrary electric and magnetic multipoles.

\section{The scalar integral family}
\label{sec:family}

Having identified the coupled tail-of-tail topology, we now
introduce the scalar integral family that underlies its
evaluation. Although motivated by gravitational radiation,
the analytical reduction is not restricted to three spatial
dimensions. We instead work in generic dimension $d$, with
a nonzero on-shell invariant, allowing the integral family
to be studied independently of a particular dimensional
expansion. This also separates the algebraic IBP identities,
which do not select a causal prescription, from the boundary
conditions and analytic continuation that determine the
integral values. We first establish the reference conventions
and their relation to gravitational radiation, before
identifying the domain of indices generated by tensor
reduction.

\subsection{Denominators, measure and kinematics}

We formulate the integral family using a Euclidean quadratic
form, with the external momentum satisfying the on-shell
condition
\begin{equation}
 k^2=m^2\ne0 .
 \label{eq:conv}
\end{equation}
Here $p$ and $q$ are $d$-dimensional loop momenta, $k$
is the external momentum, and scalar products are defined
with the Euclidean metric, so that $q^2=q\cdot q$.
The parameter $m^2$ is allowed to take complex values,
and the resulting formulas are understood as analytic
continuations of dimensionally regulated integrals.
In particular, $q^2-m^2$ is not a positive-definite
Euclidean denominator for real positive $m^2$.
A consistent choice of the branch of $\log(-m^2)$
is therefore understood throughout the calculation.

\begin{figure}[t]
\centering
\input{figures/two_point_topology.tikz}
\caption{Auxiliary two-point representation of the five
quadratic variables. Solid internal lines correspond to
$D_1$ and $D_2$, with mass parameter $m$, while dashed
lines represent massless quadratic forms. Arrows indicate
the momentum routing. The lower-left edge represents
the auxiliary ISP variable $D_5$, which appears only as
a numerator factor or is absent in the domain considered
here ($a_5\leq0$). For gravitational radiation,
$m^2=\omega^2$, with the analytic continuation specified
in eq.~\eqref{eq:causal-branches}.}
\label{fig:two-point}
\end{figure}
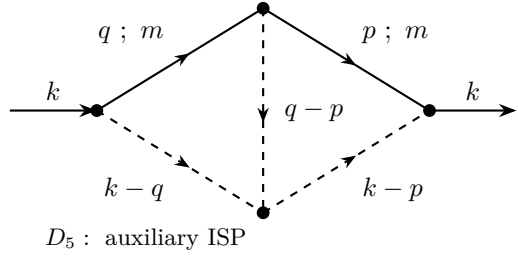

The momentum routing shown in figure~\ref{fig:two-point}
leads to five quadratic variables,
\begin{equation}
\begin{aligned}
 D_1&=q^2-m^2, &D_2&=p^2-m^2, &D_3&=(p-q)^2,\\
 D_4&=(p-k)^2, &D_5&=(q-k)^2 .
\end{aligned}
\label{eq:denoms}
\end{equation}
The first four correspond to the physical inverse
propagators of the coupled tail-of-tail diagram.
The fifth represents an irreducible scalar product
(ISP): since $q\cdot k$ cannot be expressed in
terms of $D_1,\ldots,D_4$ alone, the additional
variable $D_5$ is needed to parametrize general
polynomial numerators
\cite{Smirnov2012,Weinzierl2022}.
It does not introduce another physical propagator.

The physical denominators and polynomial numerator
insertions can then be expressed within a single
indexed family,
\begin{equation}
 I_{a_1,a_2,a_3,a_4,a_5}
 =\intq\intp
 \frac{1}{D_1^{a_1}D_2^{a_2}D_3^{a_3}D_4^{a_4}D_5^{a_5}} .
 \label{eq:defI}
\end{equation}
A vanishing index removes the corresponding
factor, while a negative integer index places
its positive power in the numerator. The mass
dimension of the integral is
$2d-2\sum_j a_j$, and the integrated two-loop
expressions carry the conventional factor
$(4\pi)^{-d}$.

The routing in figure~\ref{fig:two-point} is
consistent with momentum conservation at each
vertex. For instance, $k=q+(k-q)$ and
$q=p+(q-p)$, with the remaining momentum
relations following from the same assignments.
Dimensional regularization is understood
throughout, so scaleless integrals are set
to zero when solving the IBP relations.
Their possible separate UV and IR contributions
will be examined later, when the pole origins
are identified.

To connect this reference family with gravitational
radiation, recall that energy conservation fixes
the frequencies of the internal lines before
the loop integrations. The remaining momenta
are spatial, with $q^2=\bm q^2$ in our unbold
notation and $\bm k^2=\omega^2$. For the two
radiative propagators carrying the same frequency
$\omega$ and a common causal prescription, the
corresponding analytic continuations are
\begin{equation}
\begin{array}{c|c|c}
 &\text{radiative denominator}&\text{branch of }-m^2\\ \hline
 \text{Feynman}&\bm q^2-\omega^2-i0&-\omega^2-i0\\
 \text{retarded}&\bm q^2-(\omega+i0)^2&-(\omega+i0)^2\\
 \text{advanced}&\bm q^2-(\omega-i0)^2&-(\omega-i0)^2
\end{array}
\label{eq:causal-branches}
\end{equation}
Thus, the same analytic expressions describe
different causal boundary values once the
corresponding branch is selected. Noninteger
powers are defined by
$(-m^2)^\alpha=\exp[\alpha\log(-m^2)]$.

It is also useful to distinguish this fixed-frequency
spatial continuation from a Wick rotation of a
fully Lorentzian Feynman integral. Let
$I^{\mathrm M}_{a_1,a_2,a_3,a_4,a_5}$
denote the corresponding $d$-dimensional Lorentzian
integral with the same propagator indices and bare
loop measures, defined using the metric
$(+,-,\ldots,-)$. We denote its mass-squared
parameter by $m_{\mathrm M}^2$ and its external
momentum by $k_{\mathrm M}$, while $k_{\mathrm E}$
is the corresponding Wick-rotated Euclidean
momentum. For an integral admitting the standard
Wick rotation, the two representations are related by
\cite{Weinzierl2022}
\begin{equation}
 I^{\mathrm M}_{a_1,a_2,a_3,a_4,a_5}(m_{\mathrm M}^2)
 =i^2(-1)^{\sum_j a_j}
 I_{a_1,a_2,a_3,a_4,a_5}
 \bigl(m^2=-m_{\mathrm M}^2\bigr),
 \qquad k_{\mathrm E}^2=-k_{\mathrm M}^2,
 \label{eq:wick}
\end{equation}
where the same index tuple $(a_1,\ldots,a_5)$
is understood on both sides. Each loop rotation
contributes a factor of $i$, while the quadratic
forms change sign, producing the factor
$(-1)^{\sum_j a_j}$. The physical boundary
value must still be reached with a consistent
Feynman continuation. In particular, this
relation does not equate arbitrary Lorentzian
retarded or advanced energy contours with
Feynman integrals. For the gravitational
radiation problem considered here, no loop-energy
integration is required, and the prescriptions
are specified directly by
eq.~\eqref{eq:causal-branches}.

\subsection{The domain generated by numerator contractions}

The gravitational emission amplitudes initially
involve tensor integrals with free spatial
indices. Their decomposition into a basis built
from the external momentum and the metric,
as in a Passarino--Veltman-type reduction
\cite{PassarinoVeltman1979}, expresses the
tensor structures in terms of scalar coefficients.
After the relevant contractions, these
coefficients involve polynomials in the scalar
products of $p$, $q$, and $k$.

For the present topology, all five independent
scalar products can be written in terms of
the quadratic variables defined above:
\begin{equation}
\begin{aligned}
 q^2&=D_1+m^2, & p^2&=D_2+m^2,\\
 2p\cdot q&=D_1+D_2-D_3+2m^2,\\
 2k\cdot q&=D_1-D_5+2m^2, &
 2k\cdot p&=D_2-D_4+2m^2 .
\end{aligned}
\label{eq:products}
\end{equation}
Consequently, any polynomial numerator
generated by the tensor reduction can be
expressed in terms of $D_1,\ldots,D_5$
and $m^2$. For example, powers of $k\cdot q$
become polynomials in $D_1$, $D_5$ and
$m^2$. Since the original diagram contains
the denominator $D_1D_2D_3D_4$, these
polynomial factors can cancel propagators
or produce negative indices, but cannot
increase any of the original positive
denominator powers.

The resulting scalar integrals therefore
belong to the integer domain
\begin{equation}
 a_1,a_2,a_3,a_4\leq1,\qquad
 a_5\leq0,\qquad a_j\in\mathbb{Z} .
 \label{eq:domain}
\end{equation}
Because the numerator degree increases without
bound with the multipole rank, the negative
indices in this domain are likewise unbounded.
The exact solution developed below covers
this entire domain. It does not, however,
assume arbitrary positive propagator powers
or unrestricted complex indices.

Despite its apparent size, the domain has a
simple structure. Many integrals vanish in
dimensional regularization because they contain
scaleless subintegrations. If $a_1\leq0$,
the $q$ integration contains at most one
massless denominator, $(p-q)^2$, together
with polynomial numerator factors. After
a momentum shift, it is scaleless and
therefore vanishes. Similarly, when
$a_2\leq0$, a nonzero result requires
$a_3=a_4=1$, since canceling either
massless denominator would leave a
scaleless subintegration.

The remaining nonzero integrals naturally
fall into four sectors, classified by which
physical propagators survive. In
\emph{sector A}, one radiative denominator
is canceled, while the two massless
exchanges remain. \emph{Sector B} contains
integrals in which the connector
$(p-q)^2$ is canceled, allowing the
numerator polynomial to be expanded into
products of one-loop moments. In
\emph{sector C}, the massless denominator
$(p-k)^2$ is canceled, leaving a coupled
integral with the remaining exchange.
Finally, \emph{sector D} retains all
four physical propagators, with possible
powers of the auxiliary ISP in the numerator.

This classification organizes both the
analytical solution and the subsequent
study of the singularities. As summarized
in table~\ref{tab:map}, the sectors display
different pole structures near three
spatial dimensions, with the numerator-free
endpoint of sector D requiring separate
treatment because of its double IR pole.
The UV and IR assignments shown in the
table will be established through the
momentum-region analysis in
section~\ref{sec:poles}.

\begin{table}[t]
\centering
\small
\renewcommand{\arraystretch}{1.24}
\begin{tabular}{@{}llcll@{}}
\toprule
Sector and index pattern & Integer range & Exact in $d$ & At $d=3$ & Formula\\
\midrule
A: $I_{1,-r,1,1,-n}$ & $r,n\geq0$ & yes & simple UV & \eqref{eq:A-main}\\
B: $I_{1,1,-r,1,-n}$ & $r,n\geq0$ & yes & simple IR & \eqref{eq:B-main}\\
B: $I_{1,1,-r,-u,-n}$ & $r,u,n\geq0$ & yes & finite & \eqref{eq:B-main}\\
C: $I_{1,1,1,-r,-n}$ & $r,n\geq0$ & yes & simple UV & \eqref{eq:equality}\\
D: $I_{1,1,1,1,0}$ & endpoint & yes & double IR & \eqref{eq:D0}\\
D: $I_{1,1,1,1,-n}$ & $n\geq1$ & yes & simple IR & \eqref{eq:D-main}\\
\bottomrule
\end{tabular}
\caption{The four nonzero sectors of the integer domain.
The table lists their index patterns, allowed numerator
powers, and pole structure near $d=3$. Exact expressions
are meromorphic in $d$, with the kinematics and analytic
continuation specified in section~\ref{sec:family}.
The finite B subfamily has no dimensional pole at
$d=3$, while the double-IR endpoint of sector D
also contains a scale-dependent simple pole.
Scaleless integrals vanish in ordinary dimensional
regularization, although their separate momentum
regions may contain compensating UV and IR
contributions.}
\label{tab:map}
\end{table}

\section{IBP reduction strategy}
\label{sec:ibp}

The sector decomposition of the previous section suggests
a reduction strategy based on the simpler integrals obtained
when certain propagators are canceled. In particular, the
massless propagator $1/(p-q)^2$, which connects the two
loop momenta, will be referred to as the connector.
Removing it separates the integrations over $p$ and $q$,
whereas canceling a radiative denominator leaves a
massless bubble that can be evaluated first. These
factorized and nested configurations provide the boundary
integrals needed to solve the coupled problem. As we show
below, three integration-by-parts (IBP) identities are
sufficient to relate the remaining integrals to these
boundaries, allowing an exact solution throughout the
integer numerator domain.

\subsection{Factorized and nested boundaries}

The boundary integrals can be evaluated using standard
one-loop formulas. A particularly useful building block
is the on-shell integral
\begin{equation}
 J(a,b)=\intq\frac{1}{(q^2)^a[(q+k)^2-m^2]^b}
 =\frac{(-m^2)^{d/2-a-b}}{(4\pi)^{d/2}}
 \frac{\Gamma(a+b-d/2)\Gamma(d-2a-b)}
 {\Gamma(b)\Gamma(d-a-b)} .
 \label{eq:J}
\end{equation}
The on-shell condition $k^2=m^2$ allows this mixed
massless/radiative integral to be expressed entirely
in Gamma functions. Its derivation, together with the
standard massless bubble formula, is given in
appendix~\ref{app:oneloop}. In the applications below,
$b$ is positive, while nonpositive integer values of
$a$ correspond to polynomial numerator insertions.

The simplest boundary arises when the connector is
canceled, $a_3=0$. The two loop integrations then
factorize, giving
\begin{equation}
 I_{a,b,0,e,f}=J(f,a)J(e,b).
 \label{eq:factor}
\end{equation}
Appropriate shifts of the loop momenta bring both
factors into the form of eq.~\eqref{eq:J}. More
generally, when $a_3=-r<0$, the connector appears
in the numerator as
$[(p-q)^2]^r=(p^2+q^2-2p\cdot q)^r$.
Expanding this polynomial produces a finite sum
of products of one-loop tensor moments, whose
indices can be contracted after the integrations
have been performed separately.

A second useful boundary is obtained by canceling
one of the radiative denominators. Consider, for
example, $I_{a,0,c,e,0}$, in which the $p$
integration reduces to a massless bubble with
denominators $(p-q)^2$ and $(p-k)^2$. Evaluating
this subintegral produces a power of $(q-k)^2$,
leaving an on-shell integral of the form
$J(a,b)$ over the remaining momentum.
The resulting Gamma-function expression is
given in eq.~\eqref{eq:N} of
appendix~\ref{app:oneloop}.

Thus, both classes of boundary integrals are
known analytically in generic dimension. They
provide the starting point for the IBP reduction
of the coupled sectors, without requiring an
additional two-loop master integral with
irreducible kinematics.

\subsection{From total derivatives to closed expressions}

Having established the boundary integrals, we
now relate them to the coupled sectors through
IBP identities, obtained by integrating total
derivatives with respect to the loop momenta
\cite{Tkachov1981,ChetyrkinTkachov1981}.
For example,
\begin{equation}
 0=\intq\intp\,\frac{\partial}{\partial q^i}
 \left[(q-p)^i\prod_{j=1}^5D_j^{-a_j}\right].
 \label{eq:ibp-total}
\end{equation}
Together with the identities generated by
$(q-p)\cdot\partial_p$ and $p\cdot\partial_p$,
this gives the three index-shift relations
collected in appendix~\ref{app:recurrences}.
These relations hold between meromorphic
integrals and can be continued to the
required integer indices after introducing
suitable analytic regulation.

The reduction mechanism is illustrated by
one of their consequences:
\begin{equation}
 I_{a,1,1,e,0}
 =\frac{a}{2+a-d}
 \left[
 I_{a+1,0,1,e,0}
 -J(0,a+1)J(e,1)
 \right].
 \label{eq:boundary}
\end{equation}
For positive integer $a$ and integer $e\leq1$,
the coupled integral on the left is expressed
entirely in terms of the nested and factorized
boundaries discussed above. When $e\leq0$,
the nested contribution vanishes because its
massless subintegration is scaleless. The
relation then provides a particularly simple
starting point for the reduction of sector C.
At exceptional dimensions where an intermediate
coefficient becomes singular, the result is
understood through the limit of the complete
meromorphic expression.

An alternative way to organize the reduction
is to express every integral at fixed integer
indices as a linear combination of two seed
integrals, with coefficients rational in $d$
and involving powers of $m^2$. These seeds are
\begin{align}
 I_{1,1,0,0,0}
 &=\frac{(-m^2)^{d-2}}{(4\pi)^d}
 \Gamma(1-d/2)^2,
 \label{eq:T}\\
 I_{1,0,1,1,0}
 &=\frac{(-m^2)^{d-3}}{(4\pi)^d}
 \frac{
 \Gamma(3-d)\Gamma(2-d/2)
 \Gamma(d/2-1)^2\Gamma(2d-5)
 }{
 \Gamma(d-2)\Gamma(3d/2-3)
 } .
 \label{eq:U}
\end{align}
The first is a product of two radiative
tadpoles, while the second is the simplest
nested integral. Their distinction will
be important when identifying the momentum
origin of the dimensional poles. For the
general integer numerator domain, however,
the closed Gamma-function expressions
derived in the next section provide a more
direct representation of the solution.

\section{Exact results}
\label{sec:exact}

The four sectors identified in section~\ref{sec:family} can
be solved in an order that makes their dependence on the
simpler boundary integrals explicit. Sectors A and B are
evaluated first using nested and factorized one-loop
integrations. Their results then provide the boundary
data needed to solve the IBP recurrence for sector C.
Finally, an exact relation between sectors B and C
simplifies the reduction of sector D, in which all four
physical propagators are retained. This construction
relies on the three IBP identities
\eqref{eq:ibp1}--\eqref{eq:ibp3} and closes on the entire
integer domain of eq.~\eqref{eq:domain}. We work in
generic dimension $d$ throughout, leaving the detailed
interpretation of the dimensional poles to
section~\ref{sec:poles}.

\subsection{A: one radiative denominator canceled}

In sector A, canceling one radiative denominator leaves
a massless bubble that can be integrated before the
remaining loop. For the index pattern
$I_{1,-r,1,1,-n}$, with $r,n\geq0$, the canceled
factors appear in the numerator as $(p^2-m^2)^r$
and $(q-k)^{2n}$. Expanding these polynomials and
evaluating the massless $p$-bubble moments reduces
the problem to an on-shell one-loop integral.
The resulting expression is
\begin{align}
 I_{1,-r,1,1,-n}={}&
 \frac{(-1)^{r+n}(-m^2)^{d-3+r+n}}{(4\pi)^d}
 (d+n-2)_r(d-2)_n \nonumber\\
 &\times
 \frac{\Gamma(2-n-d/2)\Gamma(d/2-1)\Gamma(d/2+n-1)}
 {\Gamma(d+n-2)} \nonumber\\
 &\times
 \frac{\Gamma(3-r-n-d)\Gamma(2d+r+2n-5)}
 {\Gamma(3d/2+r+n-3)} .
 \label{eq:A-main}
\end{align}
Here $(x)_j=x(x+1)\cdots(x+j-1)$, with
$(x)_0=1$, denotes the Pochhammer polynomial.
This notation makes the dependence on the
integer numerator powers explicit while avoiding
ambiguous Gamma-function ratios at nonpositive
integer arguments. The result also satisfies
the IBP relations, including the recurrence
\eqref{eq:Arec} derived in
appendix~\ref{app:recurrences}.

Setting $r=n=0$ recovers the nested seed
$I_{1,0,1,1,0}$ of eq.~\eqref{eq:U}.
More generally, the factor
$(-m^2)^{d-3+r+n}$ accounts for the additional
momentum powers introduced by the numerators.
These insertions increase the short-distance
degree of the integral without introducing
new soft denominators. Accordingly, as we
will establish in section~\ref{sec:poles},
the dimensional singularity of this sector
at $d=3$ is a simple UV pole.

\subsection{B: the connector canceled}

Sector B is simpler in a different way.
With the connector $(p-q)^2$ canceled,
the two integrations factorize after expanding
the polynomial numerator. For the general
pattern $I_{1,1,-r,e,-n}$, with $r,n\geq0$
and integer $e\leq1$, the factor
$(p-q)^{2r}$ generates a finite sum of
products of one-loop moments. Combining
these contributions, or equivalently
solving the recurrence \eqref{eq:Grec}
with the factorized boundary
\eqref{eq:factor}, gives
\begin{align}
 I_{1,1,-r,e,-n}={}&
 \frac{4^n (-m^2)^{d-2+r-e+n}}{(4\pi)^d}
 \Gamma(1-r-d/2)\Gamma(1+e-d/2) \nonumber\\
 &\times\Gamma(d-1+2r)\Gamma(d-1-2e)
 \frac{\Gamma(2-r+e-d)}
 {\Gamma(2-r+e-n-d)} \nonumber\\
 &\times
 \frac{(1-r-n-d)(1+e-n-d)((d-1)/2)_n}
 {\Gamma(d+r+n)\Gamma(d-e+n)} .
 \label{eq:B-main}
\end{align}
The factorized structure is recovered
immediately at $r=0$, where
$I_{1,1,0,e,-n}=J(-n,1)J(e,1)$.
In particular, setting $r=n=e=0$
reproduces the tadpole-product seed
$I_{1,1,0,0,0}$ in eq.~\eqref{eq:T}.
For $n=0$, the result also agrees with
the corresponding one-loop-product
expression of Goldberger and Ross
\cite{GR2010}.

Although the same formula covers the
entire sector, its pole structure depends
on whether the fourth propagator survives.
For $e=1$, the remaining massless
denominator $(p-k)^2$ produces a simple
IR pole. When $e\leq0$, this denominator
is canceled, and the integral reduces
to polynomial moments of radiative
tadpoles, which are finite at $d=3$.
The two cases therefore share an
analytical solution but belong to
different classes in the UV/IR
classification developed below.

\subsection{C: a source exchange canceled}

Unlike sector B, canceling the source
exchange $(p-k)^2$ leaves the connector
$(p-q)^2$ intact, so the loop momenta
remain coupled. Nevertheless, the
corresponding IBP recurrence
\eqref{eq:Crec} can be solved using
the boundary relation \eqref{eq:boundary}
and the factorized results already obtained
for sector B. Remarkably, for arbitrary
$r,n\geq0$, the solution reduces to
the simple identity
\begin{equation}
 I_{1,1,1,-r,-n}
 =
 I_{1,1,-r,1,-n},
 \qquad r,n\geq0.
 \label{eq:equality}
\end{equation}
The right-hand side is given explicitly
by eq.~\eqref{eq:B-main} with $e=1$.
Thus, despite its coupled propagator
structure, the entire sector C is
evaluated by the same Gamma-function
expression as a factorized sector.

The equality follows from the IBP
recurrence, which lowers the numerator
degree $n$ without encountering
singular integer denominators.
At $n=0$, the boundary relation
\eqref{eq:boundary} fixes the result.
The Gamma expression for sector B
satisfies the same recurrence and
boundary condition, establishing
eq.~\eqref{eq:equality} by induction.
Appendix~\ref{app:recurrences} provides
the detailed proof, including an
alternative terminating sum and the
corresponding Gamma-function identities.
The symmetry under $r\leftrightarrow n$
also follows directly by exchanging
the loop momenta $p$ and $q$ in the
left-hand integral.

An important feature of this identity
is that equality of the regulated
integrals does not imply equality of
their pole origins. Near $d=3$, the
pole of $I_{1,1,1,-r,-n}$ is UV,
whereas that of $I_{1,1,-r,1,-n}$
is IR. Their equality holds in ordinary
dimensional regularization, where
scaleless contributions vanish and
can conceal compensating UV and IR
residues. The momentum-region analysis
of section~\ref{sec:poles} will make
this distinction explicit.

\subsection{D: all four propagators retained}

We can now turn to the fully coupled
sector D, in which all four physical
propagators remain. Its integrals
have the form $I_{1,1,1,1,-n}$,
with $n\geq0$, and contain arbitrary
powers of the irreducible numerator
$(q-k)^{2n}$. Applying the first
IBP identity \eqref{eq:ibp1} gives
the reduction \eqref{eq:D}. The
equality between sectors B and C
then produces a further simplification:
two boundary contributions cancel
exactly, leaving
\begin{equation}
 I_{1,1,1,1,-n}=
 \frac{
 I_{2,0,1,1,-n}
 -J(-n,2)J(1,1)
 }{3-n-d},
 \qquad n\geq0 .
 \label{eq:D-short}
\end{equation}
The remaining two-loop integral,
$I_{2,0,1,1,-n}$, is an auxiliary
nested boundary with a closed
Gamma-function expression given in
appendix~\ref{app:recurrences}.
Although its first index exceeds
the upper bound of the target domain,
it is needed only as a known boundary
value in the reduction.

Substituting this boundary and the
one-loop factors into eq.~\eqref{eq:D-short}
yields the exact result
\begin{align}
 I_{1,1,1,1,-n}={}&
 \frac{(-m^2)^{d+n-4}\Gamma(2-n-d/2)}
 {(4\pi)^d(3-n-d)\Gamma(d-2)}
 \Biggl[
 \frac{(-1)^n\Gamma(d/2-1)\Gamma(d/2+n-1)}
 {\Gamma(3d/2+n-4)}
 \nonumber\\
 &\hspace{8mm}\times
 \Gamma(4-n-d)\Gamma(2d+2n-6)
 \nonumber\\
 &\hspace{8mm}
 -\frac{
 \Gamma(d+2n-2)\Gamma(2-d/2)\Gamma(d-3)
 }{
 \Gamma(d+n-2)
 }
 \Biggr].
 \label{eq:D-main}
\end{align}
This expression completes the
analytical solution of the
four-propagator sector for arbitrary
integer $n\geq0$.

The singular behavior near three
spatial dimensions depends on the
numerator degree. For $n\geq1$,
the integrals contain a simple IR
pole. The endpoint $n=0$ is
different: without the additional
numerator powers, the simultaneous
soft region produces a double IR
pole. At this endpoint, the exact
result can be written more
transparently in terms of the
two seeds introduced in
section~\ref{sec:ibp}:
\begin{equation}
\begin{split}
 I_{1,1,1,1,0}={}&
 \frac{(d-2)^2}{4(d-3)^2m^4}
 I_{1,1,0,0,0}\\
 &-\frac{3d-8}{4(d-3)m^2}
 I_{1,0,1,1,0}.
\end{split}
\label{eq:D0}
\end{equation}
Although one of these seeds has
a UV pole, the complete integral
$I_{1,1,1,1,0}$ is UV convergent.
The pole origins cannot be assigned
term by term after an unseparated
IBP reduction, a point that will
be examined in detail in the
next section.

As a further check, the first
member with a nontrivial ISP
numerator satisfies
\begin{equation}
 I_{1,1,1,1,-1}=
 \frac{d-1}{2(d-3)m^2}
 I_{1,1,0,0,0}
 +\frac{5-2d}{d-2}
 I_{1,0,1,1,0}.
 \label{eq:D1}
\end{equation}
Both terms have the expected
momentum dimension, and their
sum reproduces the IR residue
obtained below. As throughout
this section, the identities
are understood as complete
meromorphic expressions in $d$;
expanding individual singular
terms before combining them
can obscure cancellations.

\section{Ultraviolet and infrared structure}
\label{sec:poles}

The exact expressions obtained in the previous section
determine the dimensional poles of the integrals, but
not the momentum regions from which they arise. This
distinction is essential in gravitational EFT: ultraviolet
(UV) divergences contribute to source-multipole
renormalization, while infrared (IR) divergences are
associated with the long-range propagation of radiation.
To identify their origins, we now specialize to
$d=3+\eps$ spatial dimensions, with
$\bm k^2=\omega^2$ and the retarded prescription.
For the dimensional expansion, we define
\begin{equation}
 \bar I_{a_1,\ldots,a_5}
 =\mu^{-2\eps}I_{a_1,\ldots,a_5},
 \qquad
 L=\gamma_E+\log\frac{-(\omega+i0)^2}{\pi\mu^2}.
 \label{eq:scale}
\end{equation}
This normalization introduces no subtraction.
Other causal boundary values follow by choosing
the corresponding branch in
eq.~\eqref{eq:causal-branches}, while the explicit
polynomial factors of $\omega^2$ remain unchanged.
Results expressed in the convention
$d=3-2\delta$ are recovered by setting
$\eps=-2\delta$, including in the pole coefficients.

The separation requires more than inspecting the
Laurent series. In particular, dimensional
regularization can conceal compensating UV and
IR contributions when scaleless integrals are
set to zero. We first illustrate this mechanism,
then construct the pole map for the complete
integer domain and establish its assignments
through an analysis of the relevant momentum
regions.

\subsection{What a scaleless zero hides}

The loss of information in scaleless integrals
can be illustrated by a simple example arising
after one loop has been integrated. Introducing
a small mass $m_{\IR}$ to regulate the infrared
region gives
\begin{align}
 \intp\frac{1}{(p^2+m_{\IR}^2)^{3-d/2}}
 &=\frac{(m_{\IR}^2)^{d-3}}{(4\pi)^{d/2}}
 \frac{\Gamma(3-d)}{\Gamma(3-d/2)}
 \nonumber\\
 &=-\frac{1}{4\pi^2\eps_{\UV}}+O(1).
 \label{eq:radialmass}
\end{align}
With the infrared singularity regulated, the
remaining pole comes from large loop momenta
and is therefore ultraviolet. If the mass is
removed from the outset, however, the integral
becomes scaleless and vanishes in dimensional
regularization. Keeping track of the two
momentum regions reveals the cancellation
hidden in this zero:
\begin{equation}
 \intp(p^2)^{-(3-d/2)}
 =-\frac{1}{4\pi^2}
 \left(
 \frac{1}{\eps_{\UV}}
 -\frac{1}{\eps_{\IR}}
 \right)=0 .
 \label{eq:scaleless}
\end{equation}
Here $\eps_{\UV}$ and $\eps_{\IR}$ label
the origins of the poles, rather than
independent dimensional regulators.
Their relative sign follows from the
convention $d=3+\eps$. The full integral
vanishes, but setting it to zero before
separating the momentum regions discards
both contributions
\cite{ManoharStewart2007,PortoRothstein2017,FS2021}.

To extract UV residues without losing
this information, we expand the integrands
in regions where the loop momenta are
large compared with the external momentum
and the radiation frequency. The radiative
denominators then admit an asymptotic
expansion in inverse powers of the hard
momenta. For example,
\begin{equation}
 \frac{1}{(p^2-\omega^2)^a}\sim
 \sum_{j\geq0}
 \frac{(a)_j}{j!}
 \frac{(\omega^2)^j}{(p^2)^{a+j}} .
 \label{eq:hard}
\end{equation}
This expansion is understood within the
hard region, with an infrared regulator
or equivalent subtraction retained when
extracting UV poles \cite{HerzogRuijl2017}.
It does not replace the original integrand
over the entire integration domain.

After angular integration, logarithmic
UV poles at $d=3$ arise only from terms
with the appropriate radial scaling.
For a fixed polynomial numerator, only
finitely many terms satisfy this condition.
Moreover, when $a=-r$ is a nonpositive
integer, the expansion in
eq.~\eqref{eq:hard} terminates.
Accounting for both the overall hard
region and possible hard subregions
therefore reduces the extraction of UV
residues to finite sums. The explicit
bounds selecting the logarithmically
divergent terms are derived in
appendix~\ref{app:regions}.

\subsection{The pole map}

Combining the exact Laurent expansions
with the residues obtained from the hard
regions allows us to determine the UV
and IR contributions throughout the
integer domain. When the hard-region
residue vanishes but the complete
integral retains a dimensional pole,
its origin must be infrared, provided
all potentially singular regions have
been accounted for. For the present
family, the factorized boundaries,
soft-region power counting and complete
hard-region sums establish this
classification, as discussed in
subsection~\ref{sec:why-poles}.

We retain separate UV and IR labels
in the following formulas to distinguish
the momentum origins of the poles.
Unless otherwise stated, $r,n\geq0$.

In sector A, where one radiative
denominator is canceled, the singular
part is
\begin{equation}
 \left.\bar I_{1,-r,1,1,-n}\right|_{\rm pole}
 =-\frac{
 (-1)^r(\omega^2)^{r+n}
 \Gamma(1+r+2n)
 }{
 64\pi^{3/2}
 \Gamma(1+n)\Gamma(3/2+r+n)
 }
 \frac{1}{\eps_{\UV}} .
 \label{eq:Apole}
\end{equation}
Thus, every nonzero member of sector A
has a simple UV pole.

The distinction between sectors B and C
is more subtle. Although their exact
regulated values coincide, their poles
originate from different momentum regions:
\begin{align}
 \left.\bar I_{1,1,-r,1,-n}\right|_{\rm pole}
 &=-\frac{
 4^{r+n}(\omega^2)^{r+n}
 }{
 32\pi^2(r+n+1)
 }
 \frac{1}{\eps_{\IR}},
 \label{eq:Bpole}\\
 \left.\bar I_{1,1,1,-r,-n}\right|_{\rm pole}
 &=-\frac{
 4^{r+n}(\omega^2)^{r+n}
 }{
 32\pi^2(r+n+1)
 }
 \frac{1}{\eps_{\UV}}.
 \label{eq:Cpole}
\end{align}
The remaining B subfamily,
$I_{1,1,-r,-u,-n}$ with
$r,u,n\geq0$, is finite at $d=3$.

Finally, in sector D, retaining all
four physical propagators gives
different singular structures
depending on the numerator degree:
\begin{align}
 \left.\bar I_{1,1,1,1,-n}\right|_{\rm pole}
 &=-\frac{
 4^n(\omega^2)^{n-1}
 }{
 128\pi^2n
 }
 \frac{1}{\eps_{\IR}},
 \qquad n\geq1,
 \label{eq:Dpole}\\
 \left.\bar I_{1,1,1,1,0}\right|_{\rm pole}
 &=-\frac{1}{128\omega^2\pi^2}
 \left(
 \frac{1}{\eps_{\IR}^2}
 +\frac{L}{\eps_{\IR}}
 \right).
 \label{eq:D0pole}
\end{align}
The numerator-free endpoint therefore
contains a double IR pole, together
with a scale-dependent simple pole
through the logarithm $L$ defined
in eq.~\eqref{eq:scale}.

These expressions complete the pole
map for the nonzero sectors of
eq.~\eqref{eq:domain}. Identifying
$\eps_{\UV}$ and $\eps_{\IR}$ with
the common dimensional regulator
recovers the ordinary Laurent
expansions, but removes the explicit
information about their momentum
origins. The complete expansions
through $O(\eps)$, including the
finite B subfamily, are collected
in appendix~\ref{app:expansion}.

\subsection{Why the pole assignments hold}
\label{sec:why-poles}

The different pole origins can be
understood from the propagator
structures of the four sectors.
For sector A, the nested integration
provides a particularly direct
argument. In $I_{1,-r,1,1,-n}$,
the massless bubble has no
logarithmic soft divergence in
three dimensions, even after
including polynomial numerator
insertions. The remaining on-shell
one-loop integral is likewise
locally integrable in its soft
region. Consequently, the pole
cannot be infrared and is
generated by the hard-momentum
region, whose expansion gives
eq.~\eqref{eq:Apole}.

The factorized structure of
sector B allows its singularities
to be examined one loop at a time.
For $I_{1,1,-r,e,-n}$, radiative
tadpole moments with integer
quadratic powers have no
logarithmic UV pole in odd
spatial dimensions. The pole
structure is therefore controlled
by the remaining massless
denominator. When $e=1$,
the corresponding on-shell
massless/radiative integral
has a simple IR pole.
For $e\leq0$, that denominator
is canceled, leaving only
finite radiative tadpole moments
at $d=3$. This explains both
branches of sector B.

Sector C provides an instructive
contrast. Although
$I_{1,1,1,-r,-n}$ remains coupled,
its only massless denominator
is $(p-q)^2$. The singularity
associated with this exchange
is locally integrable in three
dimensions under analytic
continuation of the radiative
denominators, so no logarithmic
IR pole arises. The remaining
UV residue can be followed
through the recurrence in
appendix~\ref{app:recurrences}.
At $d=3$, its factorized
inhomogeneous terms are finite,
while the homogeneous coefficient
becomes unity. The residue at
the boundary $(-r-n,0)$
therefore determines that of
the general numerator pair
$(-r,-n)$, yielding
eq.~\eqref{eq:Cpole}.

The four-propagator sector
requires a distinction between
its numerator-free endpoint
and the members with $n\geq1$.
For $I_{1,1,1,1,0}$, the overall
UV degree at $d=3$ is
$2d-8=-2$, and both one-loop
UV subgraphs are convergent.
Its dimensional poles must
therefore originate from the
infrared regions.

For $I_{1,1,1,1,-n}$ with
$n\geq1$, the simple IR pole
arises from the soft subgraph
$p\to k$. In the coefficient
of this singularity, replacing
$(p-q)^2$ by $(k-q)^2$
cancels one power of the
numerator and factorizes the
integral:
\begin{equation}
 \left.I_{1,1,1,1,-n}\right|_{\IR\ {\rm pole}}
 =
 \left.J(1-n,1)J(1,1)\right|_{\rm pole},
 \qquad n\geq1 .
 \label{eq:Dsoft}
\end{equation}
The simultaneous soft scaling
$p-k,q-k\sim\lambda$ has
degree $2n$ and is convergent
for $n\geq1$, while the second
one-loop factor introduces
no additional logarithmic pole.
The resulting IR residue
agrees with the complete
Laurent residue in
eq.~\eqref{eq:Dpole}, leaving
no UV contribution to the
dimensional pole. Higher
numerator powers may produce
power divergences, but these
do not generate logarithmic
$1/\eps$ poles at $d=3$.

\subsection{Equal integrals, different momentum regions}

The equality between sectors B
and C provides a particularly
clear example of why the
momentum-region analysis is
necessary. At $r=n=0$,
eq.~\eqref{eq:equality}
relates the factorized integral
$I_{1,1,0,1,0}$, whose pole
is infrared, to the coupled
integral $I_{1,1,1,0,0}$,
whose pole is ultraviolet.
Nevertheless, dimensional
regularization gives exactly
the same unseparated expansion
for both:
\begin{equation}
 \bar I_{1,1,0,1,0}
 =
 \bar I_{1,1,1,0,0}
 =
 -\frac{1}{32\pi^2}
 \left[
 \frac{1}{\eps}+L-1
 +\frac{\eps}{2}
 \left(
 (L-1)^2+1+\frac{\pi^2}{4}
 \right)
 \right]
 +O(\eps^2).
 \label{eq:equal-example}
\end{equation}

The equality concerns the
complete dimensionally regulated
integrals, not the individual
momentum regions of their
integrands. Their common value
is obtained after analytic
continuation and the removal
of scaleless contributions,
which can conceal compensating
UV and IR residues, as
illustrated in
eq.~\eqref{eq:scaleless}.
Keeping track of these
contributions distinguishes
the pole origins without
changing the regulated
result.

The same consideration applies
to the four-propagator endpoint.
Although its IBP reduction
in eq.~\eqref{eq:D0} involves
a UV-divergent seed, the
complete integral is UV
convergent. The individual
terms of an unseparated IBP
reduction therefore need not
retain the region assignments
of the original integral.

This distinction becomes
important when the scalar
integrals enter physical
gravitational amplitudes.
Their exact values can be
combined through ordinary
IBP reduction, but their
UV and IR contributions
must be identified separately
when extracting counterterms
or the infrared phase.
The pole map established
here provides this information
throughout the scalar integral
domain. Its sectors A--D
classify propagator-index
configurations, rather than
the momentum regions of
the physical emission diagrams.
Applying these results to
those diagrams requires
restoring their tensor
numerators and coefficients,
as we discuss in the
following section.

\section{Application to gravitational tail-of-tail amplitudes}
\label{sec:amplitudes}

Having established the exact scalar integrals and their UV/IR
structure, we now apply these results to gravitational
tail-of-tail emission amplitudes. The relevant multipolar
source couplings, bulk vertices and unintegrated electric
and magnetic amplitudes are taken from the gravitational
EFT formulation of ref.~\cite{AFS2021}, developed in
harmonic gauge. Their tensor reduction determine the
coefficients multiplying the scalar integrals studied
in the previous sections.

In the following, $s=e,m$ labels the electric and magnetic
multipole sectors, respectively, with $r\geq0$ and
$\ell=r+2$. We denote the leading emission amplitude by
$\mathcal A_0^{(s)}$ and introduce the mass-scattering
parameter $g=G_NE\omega$, where $G_N$ is Newton's constant,
$E$ the total mass-energy of the source and $\omega$ the
radiation frequency. The logarithm $L$ is defined in
eq.~\eqref{eq:scale}, and all amplitude ratios are
understood after projection onto the corresponding
tree-level tensor structure.

Among the three topologies introduced in
figure~\ref{fig:topologies}, diagrams (a) and (b)
reduce to successive one-loop integrations and
contain only UV divergences \cite{AFS2021}.
The coupled diagram (c), by contrast, contains
both UV and IR singularities and is therefore
the only topology requiring a nontrivial
amplitude-level separation. We determine its
UV and IR residues independently from the
hard- and soft-momentum regions, for electric
and magnetic multipoles of arbitrary rank.
To obtain these residues, we explicitly reduce
the tensor numerators to the scalar integral
family solved above and evaluate the resulting
contributions using their UV/IR-resolved
Laurent expansions. We present the final
results without reproducing the intermediate
tensor algebra. This calculation provides a
direct test of infrared exponentiation and,
when combined with the known UV contributions
of the remaining diagrams, of multipole
renormalization.

\subsection{Direct UV/IR separation for electric and magnetic multipoles}

The hard-region integrations yield closed
expressions for the UV residues of diagram (c)
at arbitrary multipole rank:
\begin{align}
 U_c^{(e)}(r)
 &=\frac{
 2(64r^5+552r^4+1690r^3+1905r^2+31r-1002)
 }{
 (r+2)(r+3)(2r-1)(2r+1)(2r+3)(2r+5)(2r+7)
 },
 \label{eq:Uc-direct}\\
 U_c^{(m)}(r)
 &=\frac{
 4(r+1)(4r^2+23r+36)
 }{
 (r+2)(r+3)(2r+3)(2r+5)(2r+7)
 }.
 \label{eq:Uc-magnetic}
\end{align}
The corresponding soft-region integrations
produce the same double-pole coefficient,
$-2$, for both sectors. Their simple IR
residues, however, depend on the electric
and magnetic one-loop tail coefficients
$\kappa_{r+2}$ and $\pi_{r+2}$, respectively.
In the notation of ref.~\cite{AFS2021},
these coefficients are
\begin{align}
 \kappa_{r+2}
 &=H_r+\frac{2}{r+2}
 -\frac{1}{(r+3)(r+4)},
 \label{eq:kappa-tail}\\
 \pi_{r+2}
 &=H_{r+1}
 +\frac{r+1}{(r+2)(r+3)}.
 \label{eq:pi-tail}
\end{align}
Here $H_r$ denotes the harmonic number,
with $H_0=0$.

To express both results compactly, we define
$t_\ell^{(e)}=\kappa_\ell$ and
$t_\ell^{(m)}=\pi_\ell$.
The pole structure of the coupled emission
diagram then takes the common form
\begin{equation}
 \frac{\mathcal A_c^{(s)}}{g^2\mathcal A_0^{(s)}}
 =e^{\eps L}
 \left[
 -\frac{2}{\eps_{\IR}^2}
 +\frac{4t_{r+2}^{(s)}}{\eps_{\IR}}
 +\frac{U_c^{(s)}(r)}{\eps_{\UV}}
 +O(\eps^0)
 \right],
 \qquad s=e,m.
 \label{eq:c-separated}
\end{equation}
The $\eps$ appearing in $e^{\eps L}$
is the dimensional regulator defined by
$d=3+\eps$, while the subscripts UV and IR
on the poles identify their momentum origins,
not independent regulators. The exponential
factor contains the dimensional scale dependence
and causal continuation; it is unrelated to
the auxiliary analytic regulator used to
separate the hard and soft regions. Both
this auxiliary regulator and the arbitrary
soft-region matching scale cancel from the
final residues.

Combining the independently determined UV
and IR contributions reproduces the
previously established total simple-pole
coefficients,
\begin{equation}
 U_c^{(s)}(r)+4t_{r+2}^{(s)}
 =\alpha_c^{(s)}(r),
 \qquad s=e,m,
\end{equation}
in agreement with ref.~\cite{AFS2021}.
For the quadrupoles, $r=0$, the electric
and magnetic simple IR residues are
$11/3$ and $14/3$, respectively, while
their UV residues are $334/105$ and
$8/35$. 
This agreement provides a nontrivial check of the direct
hard- and soft-region integrations, complementing the
approach of ref.~\cite{AFS2021}, where the infrared
contribution was identified through tail exponentiation
before isolating the ultraviolet pole.

\subsection{Infrared exponentiation and the universal double pole}

The direct calculation of the soft-region
contributions can now be compared with
the infrared exponentiation of the
one-loop gravitational tail. In the
same multipole normalization, the
one-loop correction is
\cite{AFS2021}
\begin{equation}
 \frac{\mathcal A_{\rm tail}^{(s)}}{\mathcal A_0^{(s)}}
 =ig\,e^{\eps L/2}
 \left[
 \frac{2}{\eps_{\IR}}
 -2t_{r+2}^{(s)}
 +O(\eps)
 \right].
 \label{eq:exponent}
\end{equation}
At order $g^2$, exponentiating this
one-loop correction produces a term
equal to one-half of its square.
The corresponding pole contribution is
\begin{equation}
 \frac12
 \left(
 \frac{\mathcal A_{\rm tail}^{(s)}}{\mathcal A_0^{(s)}}
 \right)^2
 =g^2e^{\eps L}
 \left[
 -\frac{2}{\eps_{\IR}^2}
 +\frac{4t_{r+2}^{(s)}}{\eps_{\IR}}
 +O(\eps^0)
 \right].
 \label{eq:exptest}
\end{equation}
This expression agrees exactly with
the infrared poles obtained by direct
integration of diagram (c) in
eq.~\eqref{eq:c-separated}, for electric
and magnetic multipoles of arbitrary rank.
Since diagrams (a) and (b) contain no
IR divergences, the agreement also
establishes the expected infrared
structure of the complete two-loop
emission amplitude.

The double IR pole $-2/\eps_{\IR}^2$
in eqs.~\eqref{eq:c-separated} and
\eqref{eq:exptest} is fixed by the
universal one-loop singularity
\cite{Weinberg1965,NaculichSchnitzer2011,GR2010}.
The agreement of the simple IR poles
provides an additional test, since
their coefficients depend not only on
the leading one-loop divergence but
also on the finite one-loop tail
contribution $-2t_{r+2}^{(s)}$.
Thus, the direct momentum-region
analysis verifies the complete
infrared-divergent structure
predicted by exponentiation at
the tail-of-tail order. No
$O(\eps)$ term in the one-loop
amplitude is required for this
pole-level comparison.

To make the scale dependence explicit,
we may expand the common dimensional
factor $e^{\eps L}$ in
eq.~\eqref{eq:c-separated}, obtaining
\begin{equation}
 \frac{\mathcal A_c^{(s)}}{g^2\mathcal A_0^{(s)}}
 =
 -\frac{2}{\eps_{\IR}^2}
 +\frac{4t_{r+2}^{(s)}-2L}{\eps_{\IR}}
 +\frac{U_c^{(s)}(r)}{\eps_{\UV}}
 +O(\eps^0).
 \label{eq:c-expanded}
\end{equation}
The term $-2L/\eps_{\IR}$ comes
from expanding the double IR pole
and is therefore part of the
infrared contribution, rather
than an additional UV residue.

For real nonzero frequency, the
retarded prescription gives
\begin{equation}
 L=\gamma_E+
 \log\frac{\omega^2}{\pi\mu^2}
 -i\pi\,\mathrm{sgn}(\omega).
\end{equation}
Substituting this expression into
the one-loop amplitude in
eq.~\eqref{eq:exponent} yields
\begin{equation}
 \begin{aligned}
 \frac{\mathcal A_{\rm tail}^{(s)}}{\mathcal A_0^{(s)}}
 ={}&
 ig\left[
 \frac{2}{\eps_{\IR}}
 +\gamma_E
 +\log\frac{\omega^2}{\pi\mu^2}
 -2t_{r+2}^{(s)}
 \right]\\
 &+\pi G_NE|\omega|+O(\eps).
 \end{aligned}
\end{equation}
The first term is imaginary and
contains the divergent infrared
phase, while the second is a
finite real correction to the
one-loop emission amplitude,
common to both multipole sectors.
At order $g^2$, the product of
these two contributions generates
the imaginary part of the simple
IR pole in eq.~\eqref{eq:c-expanded}.
The exponentiation test therefore
concerns the complete divergent
structure at this order, without
implying that the entire finite
tail correction is a pure phase
or determining the finite part
of the two-loop amplitude.

\subsection{Multipole renormalization and electric--magnetic consistency}

The ultraviolet divergences isolated
above contribute to the renormalization
of the source multipoles
\cite{GR2010,AFS2021}.
Let $\alpha_a^{(s)}(r)$ and
$\alpha_b^{(s)}(r)$ denote the known
UV coefficients of the two topologies
reducible to successive one-loop
integrations, using the normalization
of ref.~\cite{AFS2021} with the
common factor $g^2e^{\eps L}$.
Adding these contributions to
the independently determined
UV residue of diagram (c) gives
\begin{align}
 \beta^{(s)}(r)
 &=2\left[
 \alpha_a^{(s)}(r)
 +\alpha_b^{(s)}(r)
 +U_c^{(s)}(r)
 \right]
 \nonumber\\
 &=-\frac{
 2(15r^4+150r^3+568r^2+965r+642)
 }{
 (r+2)(r+3)(2r+3)(2r+5)(2r+7)
 },
 \qquad s=e,m.
 \label{eq:beta-electric}
\end{align}
This reproduces the established
coefficient of $g^2$ in the
source-multipole anomalous
dimension \cite{AFS2021}.
The common electric and magnetic
result is also consistent with
refs.~\cite{Fucito2025,Ivanov2025}.

For the quadrupoles, both sectors
give $\beta^{(s)}(0)=-214/105$.
More generally, the equality
between their anomalous dimensions
emerges only after the complete
UV contributions are combined.
The individual diagram-(c)
residues $U_c^{(e)}(r)$ and
$U_c^{(m)}(r)$ are different,
and their equality was never
assumed in the calculation.

The renormalization coefficients
were already known; the present
calculation provides an independent
identification of their short-distance
origin. In particular, the pole of
an individual scalar integral does
not by itself determine a source
counterterm. The latter is fixed by
the UV divergence of the assembled
emission amplitude, after combining
the contributions of the diagrams
with the coefficients determined by
their source and bulk interactions
and the corresponding tensor
contractions. The IR singularities,
in contrast, remain associated
with long-range wave propagation.

\section{Conclusions}
\label{sec:conclusions}

Gravitational tail-of-tail amplitudes involve a coupled
two-loop topology whose scalar integrals acquire increasingly
high numerator powers as the radiating multipole rank grows.
We have obtained exact closed-form expressions for the entire
relevant integer domain $a_1,\ldots,a_4\leq1$, $a_5\leq0$,
using three IBP identities and boundary integrals reducible
to standard one-loop formulas. The resulting Gamma-function
expressions solve all nonvanishing sectors in generic
dimension, while the remaining integrals vanish in
dimensional regularization. An exact relation between a
coupled and a factorized sector further simplifies the
reduction, allowing the complete four-propagator family
to be expressed in terms of known boundaries.

The generality of these results extends beyond the
gravitational problem that motivated their derivation.
The exact expressions are meromorphic in the integration
dimension and accommodate different causal prescriptions
through consistent analytic continuation. Specializing
to $d=3+\eps$, we have also derived the explicit Laurent
expansions through $O(\eps)$, including the finite
coefficients and the contributions associated with the
numerator-free double-pole endpoint. This information
is relevant for higher-order tail calculations, where
additional singular factors can make subleading
dimensional coefficients necessary. Since the solutions
are available in generic dimension, their expansions
can be extended to higher orders in $\eps$ whenever
required.

Beyond the exact integration, we have systematically
identified the ultraviolet and infrared origins of the
dimensional poles throughout the scalar integral domain
in $d=3+\eps$ spatial dimensions.
This information cannot generally be inferred from
the Laurent expansions alone, since dimensional
regularization may conceal compensating contributions
from different momentum regions. A particularly
illustrative result is the equality between two
integral sectors whose poles have opposite UV/IR
origins. The analysis also establishes the distinct
singular structures of the four-propagator sector,
including the purely infrared double pole at its
numerator-free endpoint. These results show why
the momentum-region assignments must be retained
separately when interpreting IBP-reduced integrals.

As a physical application, we have used the scalar
solutions and their UV/IR classification to determine
the divergent contributions of the coupled gravitational
tail-of-tail emission diagram for arbitrary electric
and magnetic multipoles. The independent hard- and
soft-region integrations separate its UV and IR residues,
including the overlapping soft contribution responsible
for the double pole. The complete infrared pole structure agrees with the
exponentiation of the one-loop gravitational tail,
providing an explicit, independent check of both the double and
simple poles without relying on exponentiation to
determine the UV residues. Meanwhile,
the independently determined UV residues, although
different in the electric and magnetic sectors,
reproduce the established common multipole anomalous
dimension when combined with the known contributions
of the remaining diagrams.

The exact integral solutions and their dimensional
expansions provide analytical ingredients for further
tail calculations, including higher-order radiative
processes in which additional interactions preserve
the same scalar denominator family. In particular,
they can be applied to the corresponding topologies
involving angular-momentum couplings, as well as
to suitable contributions in theories beyond general
relativity. Together with the systematic UV/IR
classification, these results provide a basis for
future calculations of gravitational amplitudes
without repeating the underlying scalar integrations.

\acknowledgments

I am grateful to Riccardo Sturani, Stefano Foffa, and Alan Müller
for many stimulating discussions over the past several years, and to Zhengwen
Liu for helpful conversations. This work was partially supported 
by the Start-up Research Fund of Southeast University under Grant No.~RF1028624160.

\appendix

\input{appendices}

\bibliographystyle{JHEP}
\bibliography{references}
\end{document}

%% file: figures/two_point_topology.tikz
\begin{tikzpicture}[x=1cm,y=1cm,
 every node/.style={font=\small},
 flow/.style={postaction={decorate},decoration={markings,mark=at position .56 with {\arrow{Stealth[length=2mm,width=1.5mm]}}}},
 massive/.style={thick,flow},massless/.style={thick,dashed,flow},
 vertex/.style={circle,fill,inner sep=1.7pt}]
 \coordinate (L) at (-2.2,0);
 \coordinate (T) at (0,1.35);
 \coordinate (R) at (2.2,0);
 \coordinate (B) at (0,-1.35);
 \draw[thick,->,>=Stealth] (-3.35,0)--node[above] {$k$}(L);
 \draw[thick,->,>=Stealth] (R)--node[above] {$k$}(3.35,0);
 \draw[massive] (L)--node[above left=2pt] {$q\ ;\ m$}(T);
 \draw[massive] (T)--node[above right=2pt] {$p\ ;\ m$}(R);
 \draw[massless] (L)--node[below left=2pt] {$k-q$}(B);
 \draw[massless] (T)--node[right=4pt] {$q-p$}(B);
 \draw[massless] (B)--node[below right=2pt] {$k-p$}(R);
 \node[font=\footnotesize] at (-1.55,-1.70) {$D_5:\ \text{auxiliary ISP}$};
 \foreach \v in {L,T,R,B} \node[vertex] at (\v) {};
\end{tikzpicture}

%% file: appendices.tex
\appendix

\section{One-loop formulas and nested boundaries}
\label{app:oneloop}
The IBP solution in the main text is built from one-loop boundary integrals. We collect their formulas here, deriving them in a region where the parameters ensure convergence and then extending the results by meromorphic continuation. All expressions use the same branch of $\log(-m^2)$, so that the boundary terms retain a consistent phase.

To derive $J(a,b)$ in eq.~\eqref{eq:J}, we combine its denominators using a Feynman parameter $x$ and shift the loop momentum by $(1-x)k$. With $k^2=m^2$, the resulting quadratic form is $q^2-(1-x)^2m^2$, and the momentum integration gives
\begin{equation}
 J(a,b)=\frac{(-m^2)^{d/2-a-b}}{(4\pi)^{d/2}}
 \frac{\Gamma(a+b-d/2)}{\Gamma(a)\Gamma(b)}
 \int_0^1\dd x\,x^{a-1}(1-x)^{d-2a-b-1}.
 \label{eq:Jparameter}
\end{equation}
The remaining beta integral yields eq.~\eqref{eq:J}. Values with nonpositive integer $a$ are understood by continuation of the integral result and correspond to polynomial numerator insertions; the parameter integral above is not evaluated separately at those indices.

The other building block is the standard massless bubble, valid for a general external momentum $k$:

\begin{equation}
 B_0(a,b;k)=\intq\frac1{(q^2)^a[(q-k)^2]^b}
 =\frac{(k^2)^{d/2-a-b}}{(4\pi)^{d/2}}
 \frac{\Gamma(a+b-d/2)\Gamma(d/2-a)\Gamma(d/2-b)}
 {\Gamma(a)\Gamma(b)\Gamma(d-a-b)}.
 \label{eq:bubble}
\end{equation}

The nested boundaries follow by performing the massless $p$-bubble integration first and applying $J$ to the remaining loop. In particular,

\begin{align}
 I_{a,0,c,e,0}&=\frac{(-m^2)^{d-a-c-e}}{(4\pi)^d}
 \frac{\Gamma(c+e-d/2)\Gamma(d/2-c)\Gamma(d/2-e)}
 {\Gamma(a)\Gamma(c)\Gamma(e)\Gamma(d-c-e)}\nonumber\\
 &\quad\times
 \frac{\Gamma(a+c+e-d)\Gamma(2d-a-2c-2e)}
 {\Gamma(3d/2-a-c-e)}.
 \label{eq:N}
\end{align}

If either $c$ or $e$ is a nonpositive integer, the massless subintegration is scaleless and must be set to zero before manipulating the Gamma factors. Otherwise, the expression is symmetric under $c\leftrightarrow e$, reflecting the exchange of the two massless bubble denominators.

\section{Detailed IBP recurrences and Gamma identities}
\label{app:recurrences}
The exact sector solutions follow from IBP identities that relate the coupled integrals to the one-loop boundaries of appendix~\ref{app:oneloop}. We first give the index-shift identities, then derive the terminating recurrence and prove the equality between sectors B and C. A few auxiliary functions simplify the repeated boundary terms in this derivation.

\subsection{Three shift identities}
Let $\boldsymbol a=(a_1,\ldots,a_5)$ and define the shift operators by $E_j^\pm I_{\boldsymbol a}=I_{\boldsymbol a\pm\boldsymbol e_j}$, where $\boldsymbol e_j$ is the unit vector in index space. Applying the total derivatives generated by $(q-p)\cdot\partial_q$, $(q-p)\cdot\partial_p$ and $p\cdot\partial_p$ gives

\begin{align}
0={}&\bigl[d-a_1-2a_3-a_5-a_1E_1^+(E_3^--E_2^-)
 +a_5E_5^+(E_4^--E_3^-)\bigr]I_{\boldsymbol a},\label{eq:ibp1}\\
0={}&\bigl[a_2+2a_3+a_4-d+a_2E_2^+(E_3^--E_1^-)
 -a_4E_4^+(E_5^--E_3^-)\bigr]I_{\boldsymbol a},\label{eq:ibp2}\\
0={}&\bigl[d-2a_2-a_3-a_4-2m^2a_2E_2^+
 +a_3E_3^+(E_1^--E_2^-)-a_4E_4^+E_2^-\bigr]I_{\boldsymbol a}.
\label{eq:ibp3}
\end{align}

The quadratic identities in eq.~\eqref{eq:products} express the differentiated products as shifts of propagator indices. In particular, eq.~\eqref{eq:ibp1} yields the boundary relation~\eqref{eq:boundary}, while combinations of all three identities generate the recurrences used below.

\subsection{The radiative-denominator boundary}
The four-propagator reduction in sector D requires an auxiliary nested boundary with a first index greater than one. For integers $a>0$, $b,f\leq0$ and $c,e\leq1$, let $S=a+b+c+e+f$. When $c=e=1$, the boundary has the exact form

\begin{align}
 I_{a,b,c,e,f}={}&\frac{(-1)^{b+f}(-m^2)^{d-S}}{(4\pi)^d}
 (d-a-2c-f+1)_{-b}(d-2c-e+1)_{-f}\nonumber\\
 &\times\frac{\Gamma(c+e+f-d/2)\Gamma(d/2-c)\Gamma(d/2-e-f)}
 {\Gamma(a)\Gamma(c)\Gamma(e)\Gamma(d-c-e-f)}\nonumber\\
 &\times\frac{\Gamma(S-d)\Gamma(2d-a-b-2c-2e-2f)}
 {\Gamma(3d/2-S)}.
 \label{eq:A}
\end{align}

If $c\leq0$ or $e\leq0$, the massless subintegration vanishes. The choice $(a,b,c,e,f)=(1,-r,1,1,-n)$ recovers sector A in eq.~\eqref{eq:A-main}, while $(2,0,1,1,-n)$ provides the auxiliary boundary entering eq.~\eqref{eq:D-short}.

The corresponding index-shift relation includes, for $a>1$ and $b<0$,

\begin{equation}
 (a-1)I_{a,b,c,e,f}
 =-(1+d-a-2c-f)I_{a-1,b+1,c,e,f},\qquad b<0.
 \label{eq:Arec}
\end{equation}

The nested integral in eq.~\eqref{eq:N} supplies the boundary at $b=f=0$. The remaining numerator powers are obtained from polynomial massless-bubble moments and the index-shift relation~\eqref{eq:Arec}.

\subsection{A terminating recurrence for sector C}
To write the recurrence for sector C compactly, introduce the auxiliary functions
\begin{equation}
 P(a,b,e,f)=J(f,a)J(e,b),\qquad C(e,f)=I_{1,1,1,e,f}.
 \label{eq:proofnotation}
\end{equation}
For $C(e,f)$, the indices satisfy $e,f\leq0$. We also write $G(c,e,f)$ for the Gamma expression in eq.~\eqref{eq:B-main}, with $r=-c$ and $n=-f$. It is used for integers $c,e\leq1$, with at least one nonpositive index, and $f\leq0$. The expression is symmetric in $c$ and $e$; at $c=0$ it reduces to the factorized boundary $P(1,1,e,f)$.

The Gamma expression satisfies the polynomial recurrence

\begin{align}
 (e-1)G(c,e,f)={}&(e-1)G(c-1,e,f+1)
 +(2c+e-d)G(c,e-1,f+1)\nonumber\\
 &+\frac{d-c-e}{2m^2}G(c-1,e-1,f+1).
 \label{eq:Grec}
\end{align}

Keeping the factor $e-1$ on the left avoids an artificial division at $e=1$. For sector C, the corresponding recurrence and its boundary value are

\begin{align}
 C(e,f)={}&\frac{2+e-d}{e-1}C(e-1,f+1)+P(1,1,e,f+1)
 +\frac{P(1,2,e-1,f+1)}{e-1},\quad f<0,\label{eq:Crec}\\
 C(e,0)={}&-\frac{P(2,1,e,0)}{3-d}.
 \label{eq:Cseed}
\end{align}

Because $e\leq0$, the denominator $e-1$ is nonzero, and each step lowers the nonnegative numerator degree $-f$. Writing $n=-f$ and $h_j=(d-e-2)_j/(1-e)_j$, iteration to the known boundary yields the finite sum

\begin{align}
 C(e,-n)={}&h_nC(e-n,0)\nonumber\\
 &+\sum_{j=0}^{n-1}h_j\left[P(1,1,e-j,1-n+j)
 +\frac{P(1,2,e-j-1,1-n+j)}{e-j-1}\right].
 \label{eq:Csum}
\end{align}

At $n=0$ the sum is empty and the boundary value is recovered. For every integer $n\geq0$, the recurrence terminates after finitely many steps, providing an alternative finite-sum expression for sector C.

\subsection{Proof of the common Gamma expression}
\label{app:proof}
The one-loop boundary in eq.~\eqref{eq:Cseed} agrees with $G(1,e,0)$. To show that the recurrence agrees as well, define $\mathcal G=G(c,e,f)$. The relevant ratios are

\begin{align}
 \frac{G(c-1,e,f+1)}{\mathcal G}
 &=\frac{(2c-d+1)(d-e-f-2)}{(d-2f-3)(c-d+e+1)},\label{eq:ratio1}\\
 \frac{G(c,e-1,f+1)}{\mathcal G}
 &=\frac{(d-2e-1)(c-d+f+2)}{(d-2f-3)(c-d+e+1)},\label{eq:ratio2}\\
 \frac{G(c-1,e-1,f+1)}{m^2\mathcal G}
 &=\frac{2(2c-d+1)(d-2e-1)(c-d+e+f+1)}
 {(c-d+e)(d-2f-3)(c-d+e+1)}.
 \label{eq:ratio3}
\end{align}

The final ratio uses $(-m^2)/m^2=-1$, with a common analytic branch for every noninteger power. Substitution into eq.~\eqref{eq:Grec} gives an identically vanishing rational expression. At $c=1$, the remaining factorized terms satisfy

\begin{equation}
 P(1,2,e-1,f+1)=\frac{d-1-e}{2m^2}P(1,1,e-1,f+1).
 \label{eq:Pratio}
\end{equation}

This identity reduces eq.~\eqref{eq:Grec} to the coupled recurrence~\eqref{eq:Crec}. Since the two expressions also share the seed, induction in $-f$ establishes
\begin{equation}
 C(e,f)=G(1,e,f),\qquad e,f\leq0 .
 \label{eq:Cgamma}
\end{equation}
The symmetry of $G$ under $c\leftrightarrow e$ then gives eq.~\eqref{eq:equality}. The proof applies to the stated integer numerator domain and does not assume unrestricted continuation in the indices.

\subsection{Cancellation in the four-propagator reduction}
The first IBP identity reduces the four-propagator integral to

\begin{equation}
 I_{1,1,1,1,f}=\frac{fC(0,f+1)+I_{2,0,1,1,f}-P(2,1,1,f)-fP(1,1,1,f+1)}{3+f-d}.
 \label{eq:D}
\end{equation}

For $f<0$, the sector B/C identity gives
\begin{equation}
 C(0,f+1)=G(1,0,f+1)=G(0,1,f+1)=P(1,1,1,f+1).
 \label{eq:Dcancel}
\end{equation}
The contributions proportional to $f$ consequently cancel. At $f=0$, they vanish before their arguments are evaluated, so the out-of-domain quantity $C(0,1)$ is never needed. In both cases, we recover eq.~\eqref{eq:D-short} with $n=-f\geq0$. Substituting the boundary formulas~\eqref{eq:J} and \eqref{eq:A} gives eq.~\eqref{eq:D-main}. At $n=0$, the result can also be expressed in terms of the two seed integrals $I_{1,1,0,0,0}$ and $I_{1,0,1,1,0}$, as shown in eq.~\eqref{eq:D0}.

\subsection{Relation to the quadrupole integral boundaries}
The normalization can also be checked against the quadrupole calculation of Goldberger and Ross~\cite{GR2010}. With the momentum change $q=p_{\rm GR}+l_{\rm GR}$ and $p=l_{\rm GR}$, their appendix-A family is related to ours by
\begin{equation}
 I_2(n_1,n_3)=I_{1,1,n_3,n_1,0}.
 \label{eq:GRmap}
\end{equation}
The spatial dimension agrees with our $d$, and the Feynman boundary value used there coincides with the retarded prescription at positive frequency. Our boundary relation~\eqref{eq:boundary} reproduces their eq.~(A13), while eq.~\eqref{eq:B-main} at $n=0$ agrees with their eq.~(A14). These comparisons check the normalization independently of the arbitrary-numerator recurrences.

\section{Laurent expansions through linear order}
\label{app:expansion}
We now specialize the exact solutions to $d=3+\eps$ with the retarded prescription and the normalized integrals $\bar I_{\boldsymbol a}$ and logarithm $L$ of eq.~\eqref{eq:scale}. The expansions include every coefficient through $O(\eps)$, with remainder $O(\eps^2)$. The UV and IR origins were established in section~\ref{sec:poles}; here a single unlabelled $\eps$ denotes the dimensional regulator. We use the generalized harmonic numbers
\begin{equation}
 H_j^{(a)}=\sum_{l=1}^j\frac1{l^a},\qquad H_j=H_j^{(1)},\qquad H_0^{(a)}=0.
 \label{eq:harmonic}
\end{equation}
For each simple-pole sector, $F$ denotes the finite coefficient of the logarithm of the pole-subtracted integral, while $Q$ collects its quadratic-order logarithmic contribution. This notation keeps the formulas compact in terms of harmonic sums.

\subsection{The simple-pole sectors}
For sector A, with $r,n\geq0$, the Laurent expansion is
\begin{align}
 \bar I_{1,-r,1,1,-n}={}&
 -\frac{(-1)^r(\omega^2)^{r+n}\Gamma(1+r+2n)}
 {64\pi^{3/2}\Gamma(1+n)\Gamma(3/2+r+n)}\nonumber\\
 &\times\left[\frac1\eps+F_A+\frac\eps2(F_A^2+Q_A)\right]+O(\eps^2),
 \label{eq:Aexp}\\
 F_A={}&L-H_n-3H_{2r+2n+2}+\frac32H_{r+n+1}+2H_{r+2n},
 \label{eq:FA}\\
 Q_A={}&H_n^{(2)}+9H_{2r+2n+2}^{(2)}-\frac94H_{r+n+1}^{(2)}
 -4H_{r+2n}^{(2)}-\frac{\pi^2}{12}.
 \label{eq:QA}
\end{align}
At $r=n=0$, one obtains $F_A=L-3$ and $Q_A=9-\pi^2/12$, in agreement with the nested seed. The different dependence on $r$ and $n$ reflects the distinct numerator insertions.

The IR-singular branch of sector B ($a_4=1$) and sector C share an exact value. Their common Laurent expansion is
\begin{align}
 \bar I_{1,1,-r,1,-n}={}&\bar I_{1,1,1,-r,-n}\nonumber\\
 ={}&-\frac{4^{r+n}(\omega^2)^{r+n}}{32\pi^2(r+n+1)}
 \left[\frac1\eps+F_{BC}+\frac\eps2(F_{BC}^2+Q_{BC})\right]+O(\eps^2),
 \label{eq:BCexp}\\
 F_{BC}={}&L-\frac1{r+n+1}-\frac{H_r+H_n}{2},
 \label{eq:FBC}\\
 Q_{BC}={}&\frac{\pi^2}{4}+\frac34\left(H_r^{(2)}+H_n^{(2)}\right)
 +\frac1{(r+n+1)^2}.
 \label{eq:QBC}
\end{align}
Although the expansions agree term by term, the pole belongs to the IR region in sector B and to the UV region in sector C.

For sector D with $n\geq1$, we find
\begin{align}
 \bar I_{1,1,1,1,-n}={}&-\frac{4^n(\omega^2)^{n-1}}{128n\pi^2}
 \left[\frac1\eps+F_D+\frac\eps2(F_D^2+Q_D)\right]+O(\eps^2),
 \label{eq:Dexp}\\
 F_D={}&L+H_{2n}-\frac32H_n-\frac1n,
 \label{eq:FD}\\
 Q_D={}&\frac{7\pi^2}{12}-5H_{2n}^{(2)}+\frac{11}{4}H_n^{(2)}
 +\frac1{n^2}-2(H_{2n}-H_n)^2.
 \label{eq:QD}
\end{align}
Unlike the preceding sectors, the exact result for D is a difference of two Gamma products. Expanding these products before combining them is necessary to recover the last squared term in $Q_D$.

\subsection{The finite B subfamily and scaleless sectors}
When the fourth propagator is also canceled, sector B is finite at $d=3$. For $I_{1,1,-r,-u,-n}$ with $r,u,n\geq0$ and $t=r+u+n$, its expansion reads
\begin{align}
 \bar I_{1,1,-r,-u,-n}={}&
 -\frac{4^{t-2}(\omega^2)^{t+1}r!\,u!\,n!\,(t+1)!}
 {\pi^2(r+u+1)!(r+n+1)!(u+n+1)!}\nonumber\\
 &\times\left[1+\eps F_{\rm fin}\right]+O(\eps^2),
 \label{eq:Bfinite}\\
 F_{\rm fin}={}&L+H_{t+1}-H_{r+u+1}-H_{r+n+1}-H_{u+n+1}
 +\frac{H_r+H_u+H_n}{2}.
 \label{eq:Ffin}
\end{align}
The formula covers three independent numerator degrees. Setting $r=u=0$ gives $-4^{n-2}(\omega^2)^{n+1}/[(n+1)\pi^2]$ at leading order, while $r=u=n=0$ recovers the tadpole-product seed. The remaining scaleless sectors vanish identically in dimensional regularization, so all their unseparated Laurent coefficients are zero; this does not assign separate values to their UV and IR regions.

\subsection{The double-pole endpoint and seed checks}
The numerator-free endpoint of sector D has a double pole and must be expanded separately from the $n\geq1$ family:
\begin{align}
 \bar I_{1,1,1,1,0}={}&-\frac1{128\omega^2\pi^2}
 \Biggl[\frac1{\eps^2}+\frac L\eps+\frac{L^2}{2}+\frac{7\pi^2}{24}
 \nonumber\\
 &\hspace{13mm}+\eps\left(\frac{L^3}{6}
 +\frac{7\pi^2L}{24}-\frac{53\zeta(3)}{12}\right)\Biggr]+O(\eps^2).
 \label{eq:D0exp}
\end{align}
The simple-pole expression~\eqref{eq:Dexp} cannot be specialized to $n=0$. For comparison with the exact endpoint reduction, the two seed expansions are
\begin{align}
 \bar I_{1,1,0,0,0}&=-\frac{\omega^2}{16\pi^2}
 \left[1+\eps(L-2)\right]+O(\eps^2),
 \label{eq:Texp}\\
 \bar I_{1,0,1,1,0}&=-\frac1{32\pi^2}
 \left[\frac1\eps+L-3+\frac\eps2
 \left((L-3)^2+9-\frac{\pi^2}{12}\right)\right]+O(\eps^2).
 \label{eq:Uexp}
\end{align}
The two-seed identity~\eqref{eq:D0} provides an independent check. Its rational coefficients contain $1/\eps^2$ and $1/\eps$, so reproducing the endpoint through $O(\eps)$ requires the finite seed through $O(\eps^3)$ and the nested seed through $O(\eps^2)$. Expanding the exact seed expressions to these orders reproduces the $\zeta(3)$ contribution in eq.~\eqref{eq:D0exp}.

\subsection{Derivation, validation and use in higher iterations}
For a representative simple-pole integral $\bar I_{\boldsymbol a}$ with residue $R_{\boldsymbol a}$, the regular part can be expanded logarithmically:
\begin{equation}
 \frac{\eps\bar I_{\boldsymbol a}}{R_{\boldsymbol a}}
 =\exp\left[\eps F+\frac{\eps^2}{2}Q+O(\eps^3)\right].
 \label{eq:logexp}
\end{equation}
Expanding the Gamma functions near $d=3$ produces derivatives of $\log\Gamma$ at integer and half-integer arguments. These can be expressed in terms of the harmonic sums introduced above and known constants. For sector D, the two Gamma products must be expanded separately before subtraction; treating their difference as a single product would omit the last term of eq.~\eqref{eq:QD}. At the double-pole endpoint, third-order derivatives also generate the $\zeta(3)$ term.

The analytical expansions were checked by substituting representative integer indices into the exact formulas and expanding independently. The implementation compares 227 coefficients at 84 index points, including the finite recurrence for sector C and a two-seed check of the D endpoint. Numerical comparisons at 90-digit precision further test a point away from the branch cut and both physical cut boundaries. All 252 branch cases show the expected $O(\eps^2)$ remainder. These checks support, but do not replace, the general Gamma-function derivation.

The $O(\eps)$ coefficients become relevant when these integrals multiply a further simple pole. More generally, a factor with a pole of order $p$ may require the accompanying expansion through $O(\eps^p)$ to determine a finite product, and higher orders for subleading contributions. The required order therefore depends on the structure of each higher-tail or tail-dressed-memory calculation. Since the underlying formulas are exact in $d$, the expansions can be extended accordingly.

\section{Finite sums for the ultraviolet regions}
\label{app:regions}
The UV analysis of section~\ref{sec:poles} starts from hard-region expansions of the radiative denominators and polynomial numerators. After one loop is integrated, the remaining terms are massless one-loop expressions. Their UV residues must be extracted with care: canceling a denominator can produce a scaleless integral whose vanishing hides a logarithmic UV contribution. We therefore distinguish the cases in which a massless denominator survives from those in which only a polynomial remains, and show how both lead to finite residue sums.

\subsection{Polynomial bubble moments}
The first step is to evaluate the contracted polynomial moments of a massless bubble. With $\mathcal B$ denoting the corresponding Gamma-function coefficient, we use
\begin{align}
 \intq\frac{(k\cdot q)^s}{(q^2)^a[(p-q)^2]^b}
 ={}&\frac{(p^2)^{d/2-a-b}}{(4\pi)^{d/2}}
 \sum_{r=0}^{\lfloor s/2\rfloor}
 \frac{s!(p^2\omega^2)^r(k\cdot p)^{s-2r}}
 {4^rr!(s-2r)!}\,
 \mathcal B(a,b;r,s),
 \label{eq:scalarbubble}\\
 \mathcal B(a,b;r,s)={}&
 \frac{\Gamma(a+b-r-d/2)\Gamma(d/2+s-a-r)\Gamma(d/2-b+r)}
 {\Gamma(a)\Gamma(b)\Gamma(d+s-a-b)}.
 \label{eq:Bcoef}
\end{align}
Here $\lfloor x\rfloor$ denotes the greatest integer not exceeding $x$, and $r$ counts the pairs of indices contracted in the rotationally invariant loop integration. At $s=0$, the expression reduces to the ordinary massless bubble. Only scalar contractions are needed for the subsequent residue sums.

\subsection{The residual integral and its two branches}
Consider $I_{1,a_2,a_3,a_4,-n}$ with $n\geq0$. The integers $i,j$ label the hard expansions of $D_1$ and $D_2$, while $l,s,r,\alpha,\beta$ index the finite polynomial expansions that follow. Their ranges are $0\leq l\leq n$, $0\leq s\leq l$, $0\leq r\leq\lfloor s/2\rfloor$, $0\leq\alpha\leq s-2r$ and $0\leq\beta\leq\alpha$. After the first integration, the remaining massless $p$ bubble has indices
\begin{equation}
 A_p=H-\frac d2,\qquad b=a_4-\beta,\qquad
 H=i+j+1+a_2+a_3+s-l-r-\alpha+\beta .
 \label{eq:Hregion}
\end{equation}
The symbol $H$ abbreviates the indicated combination of summation indices and is used only in this appendix. Whether the last bubble contains a massless denominator is determined by the sign of $b$:
\begin{align}
 b>0:&\qquad 0\leq\beta\leq\min(\alpha,a_4-1),
 \label{eq:betapos}\\
 b\leq0:&\qquad \max(0,a_4)\leq\beta\leq\alpha.
 \label{eq:betazero}
\end{align}
For $b>0$ the massless denominator survives; for $b\leq0$ it is canceled and leaves a polynomial. In the domain of eq.~\eqref{eq:domain}, the first case requires $a_4=1$ and $\beta=0$.
The coefficient of each residual bubble is
\begin{align}
 W_{ijlsr\alpha\beta}={}&
 \binom nl\binom ls\binom{s-2r}{\alpha}\binom\alpha\beta
 \frac{(a_2)_j}{j!}
 \frac{(-1)^{s+\beta}s!}{r!(s-2r)!}\nonumber\\
 &\times\frac{(\omega^2)^{i+j+n+s-r-l-\alpha}}{(4\pi)^{d/2}}
 \mathcal B(1+i+s-l,a_3;r,s).
 \label{eq:W}
\end{align}
The UV residue is extracted from the regulated hard-region sum
\begin{equation}
 \sum_{i,j\geq0}\sum_{l=0}^n\sum_{s=0}^l
 \sum_{r=0}^{\lfloor s/2\rfloor}\sum_{\alpha=0}^{s-2r}
 \sum_{\beta=0}^{\alpha}
 W_{ijlsr\alpha\beta}\,B_0(H-d/2,a_4-\beta;k),
 \label{eq:regionsum}
\end{equation}
with $H$ defined in eq.~\eqref{eq:Hregion}. Its UV residue is extracted only after regulating the complementary momentum regions; the sum is not a representation of the full integral's infrared behavior.

\subsection{The non-scaleless branch}
When $b=a_4-\beta>0$, the surviving denominator restricts the summation range through eq.~\eqref{eq:betapos}. At $d=3$, possible UV terms satisfy $H+b\leq3$, whereas an IR singularity at $p=0$ would require $H\geq3$. Since $b=1$ in the target domain, these conditions do not overlap, and the UV bound becomes
\begin{equation}
 0\leq j\leq 2-a_2-a_3-a_4-s+l+r+\alpha-i.
 \label{eq:jpos}
\end{equation}
Terms with a negative upper limit are absent. When $a_3=1$, the nonvanishing coefficients in eq.~\eqref{eq:Bcoef} impose the additional restriction
\begin{equation}
 \max(0,l-s)\leq i\leq l.
 \label{eq:ibounds}
\end{equation}
This restriction follows from zeros of reciprocal Gamma functions at integer arguments, while the numerator Gamma functions remain at half-integer arguments. For $a_3\leq0$, the factorized polynomial representation is more direct.

\subsection{The scaleless branch}
If $b\leq0$, the last massless denominator has been canceled. Under the bounds in eq.~\eqref{eq:betazero}, expand the resulting polynomial with
\begin{equation}
 0\leq t\leq\beta-a_4,\qquad u=2v,\qquad
 0\leq v\leq\lfloor t/2\rfloor.
 \label{eq:tvbounds}
\end{equation}
Only even angular powers survive, hence $u=2v$. A logarithmic UV term must satisfy
\begin{equation}
 i+j+1+a_2+a_3+s-l-r-\alpha+\beta-t+v=3.
 \label{eq:logcondition}
\end{equation}
This condition fixes $j$ and removes the apparent infinite sum:
\begin{equation}
 j=2-a_2-a_3-s+l+r+\alpha-\beta+t-v-i\geq0.
 \label{eq:jzero}
\end{equation}
The allowed range of $i$ is therefore bounded above by the right-hand side of eq.~\eqref{eq:jzero} with its $-i$ term removed, and must also satisfy eq.~\eqref{eq:ibounds} when $a_3=1$. For nonpositive integer $a_2$, the factor $(a_2)_j$ further restricts $j\leq-a_2$.

Each surviving term contributes $W_{ijlsr\alpha\beta}$ multiplied by
\begin{equation}
 \binom{\beta-a_4}{t}\binom t{2v}4^v
 (\omega^2)^{\beta-a_4-t+v}
 \frac{(2v-1)!!\Gamma(d/2)}{2^v\Gamma(d/2+v)}
 \left(-\frac1{4\pi^2\eps_{\UV}}\right).
 \label{eq:zeroterm}
\end{equation}
The remaining prefactor is evaluated at $d=3$. We use $(-1)!!=1$, and the angular factor reduces to $1/(2v+1)$. Together, eqs.~\eqref{eq:W}, \eqref{eq:betazero} and \eqref{eq:tvbounds}--\eqref{eq:zeroterm} specify the finite set of terms contributing to the UV residue; an empty sum contributes zero.

Two examples check the separation. For the sector-A seed $I_{1,0,1,1,0}$, only the non-scaleless branch contributes, giving $-1/(32\pi^2\eps_{\UV})$. For $I_{1,1,1,1,-1}$, the two branches yield $+1/(64\pi^2\eps_{\UV})$ and $-1/(64\pi^2\eps_{\UV})$, which cancel. The vanishing UV residue agrees with the soft-region classification in section~\ref{sec:poles}.